\documentclass[a4paper,fleqn,usenatbib]{mnras}

\usepackage[T1]{fontenc}
\usepackage{ae,aecompl}

\usepackage{graphicx}	
\usepackage{amssymb}	
\usepackage{amsmath}
\usepackage{bm}
\usepackage{multirow}
\usepackage{mn2e-breakabs}

\def\dD{\delta_{\rm D}}

\newcommand{\beq}{\begin{equation}}
\newcommand{\eeq}{\end{equation}}
\newcommand{\beqa}{\begin{eqnarray}}
\newcommand{\eeqa}{\end{eqnarray}}

\font\BF=cmmib10

\def\kk{{\hbox{\BF k}}}
\def\x{{\hbox{\BF x}}}
\def\r{{\hbox{\BF r}}}

\def\q{{\hbox{\BF q}}}

\title[Anisotropic clustering in the completed BOSS]{
The clustering of galaxies in the completed SDSS-III Baryon Oscillation
 Spectroscopic Survey: cosmological implications of the configuration-space
 clustering wedges 
}

\author[A. G. S\'anchez et al.]{Ariel G. S\'anchez,$^{1}$\thanks{E-mail: arielsan@mpe.mpg.de}
Rom\'an Scoccimarro$^{2}$,
Mart\'{\i}n Crocce$^{3}$,
Jan~Niklas Grieb$^{4,1}$,
\newauthor  Salvador Salazar-Albornoz$^{4,1}$,
Claudio DallaVecchia$^{5,6}$,
Martha Lippich$^{4,1}$,
\newauthor 
Florian Beutler$^{7,8}$,
Joel~R. Brownstein$^{9}$,
Chia-Hsun Chuang$^{10,11}$,
Daniel~J. Eisenstein$^{12}$,
\newauthor Francisco-Shu Kitaura$^{11,13,8}$,
Matthew~D. Olmstead$^{14}$,
Will~J. Percival$^{7}$,
\newauthor Francisco Prada$^{10,15,16}$,
 Sergio Rodr\'{\i}guez-Torres$^{10,15,17}$,
 Ashley J. Ross$^{18,7}$,
\newauthor Lado Samushia$^{19,20,7}$, 
Hee-Jong Seo$^{21}$,
Jeremy Tinker$^{2}$, 
Rita Tojeiro$^{22}$, 
\newauthor Mariana Vargas-Maga\~na$^{23,24,25}$, 
Yuting Wang$^{26,7}$ \&
Gong-Bo Zhao$^{26,7}$
\\
$^{1}$ Max-Planck-Institut f\"ur extraterrestrische Physik, Postfach 1312, Giessenbachstr., 85741 Garching, Germany\\ 
$^{2}$ Center for Cosmology and Particle Physics, Department of Physics, New York University, NY 10003, New York, USA\\ 
$^{3}$ Institut de Ci\`encies de l'Espai, IEEC-CSIC, Campus UAB, Carrer de Can Magrans, s/n, 08193 Bellaterra, Barcelona, Spain\\ 
$^{4}$ Universit\"ats-Sternwarte M\"unchen, Ludwig-Maximilians-Universit\"at M\"unchen, Scheinerstrasse 1, 81679 Munich, Germany \\
$^{5}$ Instituto de Astrof\'\i{}sica de Canarias, C/ V\'\i{}a L\'actea s/n, 38205 La Laguna, Tenerife, Spain\\
$^{6}$ Departamento de Astrof\'\i{}sica, Universidad de La Laguna, Av{.} del Astrof\'\i{}sico Francisco S\'anchez s/n, 38206 La Laguna, Tenerife, Spain\\
$^{7}$ Institute of Cosmology \& Gravitation, Dennis Sciama Building, University of Portsmouth, Portsmouth, PO1 3FX, UK\\
$^{8}$ Lawrence Berkeley National Laboratory, 1 Cyclotron Roadd, Berkeley CA 94720, USA\\
$^{9}$ Department of Physics and Astronomy, University of Utah, 115 S 1400 E, Salt Lake City, UT 84112, USA \\
$^{10}$ Instituto de F\'{\i}sica Te\'orica, (UAM/CSIC), Universidad Aut\'onoma de Madrid, Cantoblanco, E-28049 Madrid, Spain \\
$^{11}$ Leibniz-Institut f\"ur Astrophysik Potsdam (AIP), An der Sternwarte 16, D-14482 Potsdam, Germany \\
$^{12}$ Harvard-Smithsonian Center for Astrophysics, 60 Garden St., Cambridge, MA 02138, USA \\
$^{13}$ Departments of Physics and Astronomy, University of California, Berkeley, CA 94720, USA \\
$^{14}$ Department of Chemistry and Physics, King's College, 133 North River St, Wilkes Barre, PA 18711, USA\\
$^{15}$ Campus of International Excellence UAM+CSIC, Cantoblanco, E-28049 Madrid, Spain\\
$^{16}$ Instituto de Astrof\'{\i}sica de Andaluc\'{\i}a (CSIC), Glorieta de la Astronom\'{\i}a, E-18080 Granada, Spain\\
$^{17}$ Departamento de F\'{\i}sica Te\'orica, Universidad Aut\'onoma de Madrid, Cantoblanco, 28049, Madrid, Spain\\
$^{18}$ Center for Cosmology and Astro-Particle Physics, Ohio State University, Columbus, OH 43210, USA\\
$^{19}$ Kansas State University, Manhattan KS 66506, USA\\
$^{20}$ National Abastumani Astrophysical Observatory, Ilia State University, 2A Kazbegi Ave., GE-1060 Tbilisi, Georgia\\
$^{21}$ Department of Physics and Astronomy, Ohio University, 251B Clippinger Labs, Athens, OH 45701, USA\\
$^{22}$ School of Physics and Astronomy, University of St Andrews, North Haugh, St Andrews KY16 9SS, UK\\
$^{23}$ Instituto de F\'isica, Universidad Nacional Aut\'onoma de M\'exico, Apdo. Postal 20-364, M\'exico\\
$^{24}$ Department of Physics, Carnegie Mellon University, 5000 Forbes Ave., Pittsburgh, PA 15217, USA\\
$^{25}$ McWilliams Center for Cosmology, Carnegie Mellon University, 5000 Forbes Ave., Pittsburgh, PA 15217, USA\\
$^{26}$ National Astronomy Observatories, Chinese Academy of Science, Beijing, 100012, P.R.China
}

\date{Submitted to MNRAS}

\pubyear{2016}

\begin{document}
\label{firstpage}
\pagerange{\pageref{firstpage}--\pageref{lastpage}}
\maketitle

\begin{abstract}
We explore the cosmological implications of anisotropic clustering measurements in configuration space of
the final galaxy samples from Data Release 12 of the SDSS-III Baryon Oscillation Spectroscopic Survey. 
We implement a new detailed modelling of the effects of non-linearities, galaxy bias and 
redshift-space distortions that can be used to extract unbiased cosmological information from our measurements 
for scales $s \gtrsim 20\,h^{-1}{\rm Mpc}$.
We combined the galaxy clustering information from BOSS with the latest cosmic microwave
background (CMB) observations and Type Ia supernovae samples 
and found no significant evidence for a deviation from the $\Lambda$CDM cosmological model.
In particular, these data sets can constrain the dark 
energy equation of state parameter to $w_{\rm DE}=-0.996\pm0.042$ when assumed time-independent, 
the curvature of the Universe to $\Omega_{k}=-0.0007\pm 0.0030$ and the sum of the neutrino masses to 
$\sum m_{\nu} < 0.25\,{\rm eV}$ at 95 per cent CL.
We explore the constraints on the growth rate of cosmic structures assuming $f(z)=\Omega_{\rm m}(z)^\gamma$
and obtain $\gamma = 0.609\pm 0.079$, in good agreement with the predictions of general relativity of $\gamma=0.55$.
We compress the information of our clustering measurements into constraints on
the parameter combinations $D_{\rm V}(z)/r_{\rm d}$, $F_{\rm AP}(z)$ and $f\sigma_8(z)$ 
at the effective redshifts of $z=0.38$, 0.51 and 0.61 with their respective 
covariance matrices and find good agreement with the predictions for these parameters obtained from 
the best-fitting $\Lambda$CDM model to the CMB data from the Planck satellite. 
This paper is part of a set that analyses the final galaxy clustering dataset from BOSS.
The measurements and likelihoods presented here are combined with others in \citet{Acacia2016} to 
produce the final cosmological constraints from BOSS.
\end{abstract}




\section{Introduction}
\label{sec:intro}

Measurements of the large-scale clustering of galaxies offer a  
powerful route to obtain accurate cosmological information
\citep{Davis1983,Maddox1990,Tegmark2004,Cole2005,Eisenstein2005,Anderson2012,Anderson2013,Anderson2014}.
Two-point statistics such as the power spectrum, $P(k)$, and its Fourier transform, the two-point
correlation function $\xi(s)$, have been the preferred tools for analyses of the large-scale structure (LSS) 
of the Universe. The shape of these measurements can be used to constrain the values of several cosmological parameters, providing clues about the nature of dark energy, potential deviations from the predictions
of general relativity (GR), the physics of inflation, neutrino masses, etc. \citep{Percival2002,Percival2010,Tegmark2004,Sanchez2006,Sanchez2009,Sanchez2012,Blake2011,
Parkinson2012}.

A particularly important source of cosmological information contained in the large-scale galaxy clustering
pattern is the signature of the baryon acoustic oscillations (BAO), 
which are the vestige of acoustic waves that propagated through the photon-baryon fluid prior to
recombination. The BAO signature was first detected by \citet{Eisenstein2005} in the correlation function
of the luminous red galaxy sample of the Sloan Digital Sky Survey \citep[SDSS,][]{York2000}, 
where it can be seen as a broad peak on large scales \citep{Matsubara2004}, 
and by \citet{Cole2005} in the power spectrum of the Two-degree Field Galaxy Redshift survey \citep[2dFGRS,][]{Colless2001,Colless2003}, where it appears as a series of wiggles \citep{Eisenstein1998,Meiksin1999}. 
The position of the peak in the correlation function and the wavelength of the oscillations in the power
spectrum  closely match the sound horizon scale at the drag redshift, 
$r_{\rm d}\simeq150\,{\rm Mpc}$. 
This means that the BAO scale inferred from the clustering of galaxies in the directions
parallel and perpendicular to the line of sight can be used as a standard ruler
to measure the Hubble parameter, $H(z)$, and the angular diameter distance, $D_{\rm M}(z)$,
through the Alcock--Paczynski (AP) test \citep{Alcock1979,Blake2003,Linder2003}.

As the AP test cannot be applied to angle-averaged clustering measurements, the full power 
of the BAO signal can only be exploited by means of anisotropic clustering measurements. 
That means measurements of the full two-dimensional correlation
function or power spectrum \citep{Wagner2008,Shoji2009}, their Legendre
multipole moments \citep{Padmanabhan2008} or the clustering wedges statistic \citep*{Kazin2012}.
These measurements are affected by redshift-space distortions (RSD) due to the peculiar velocities
of the galaxies along the line of sight, which are significantly larger than the geometric 
distortions due to the AP effect and must be accurately modelled to avoid introducing systematic errors 
in the obtained constraints. 
However, more than a complication for the application of the AP test, RSD provide 
additional cosmological information, as they can be used to constrain the growth rate of cosmic
structures \citep{Guzzo2008}. In this way, thanks to the joint information from BAO and RSD, 
anisotropic clustering measurements can provide information on the expansion history of the 
Universe and the growth rate of density fluctuations, which is essential to distinguish between
dark energy and modified gravity as the driver of cosmic acceleration. 

Previous analyses of anisotropic clustering measurements based on data from the SDSS-III 
\citep{Eisenstein2011} Baryon Oscillation Spectroscopic Survey \citep[BOSS,][]{Dawson2013}, 
clearly illustrated their constraining power \citep{Anderson2013,Anderson2014,Reid2012,Samushia2013,Samushia2014,Chuang2013a,
Beutler2014}. In particular, \citet{Sanchez2013,Sanchez2014} explored the cosmological implications 
of the full shape of measurements of two clustering wedges based on the galaxy samples of BOSS DR11.
In this paper we extend these analyses to the final galaxy samples from BOSS, corresponding to 
SDSS DR12 \citep{Alam2015}. The volume probed by DR12 is only $\sim10$ per cent larger than that of DR11.
For this reason, we focus on improving our analysis methodology in order to 
maximize the cosmological information extracted from the sample. 
We make use of the joint information of the LOWZ and CMASS galaxy samples into the {\it combined} BOSS sample 
described in \citet{Reid2016}, increasing the effective volume of the survey
with respect to the separate analysis of these samples \citep{Acacia2016}.
We also use state-of-the-art models of the effect of non-linearities, bias and redshift-space
distortions that allow us to extend our analysis of the full shape of the clustering wedges to smaller scales. 
We perform extensive tests of the performance of our methodology on $N$-body simulations and 
mock catalogues and find precise and accurate constraints.

\begin{figure*}
\includegraphics[width=0.98\textwidth]{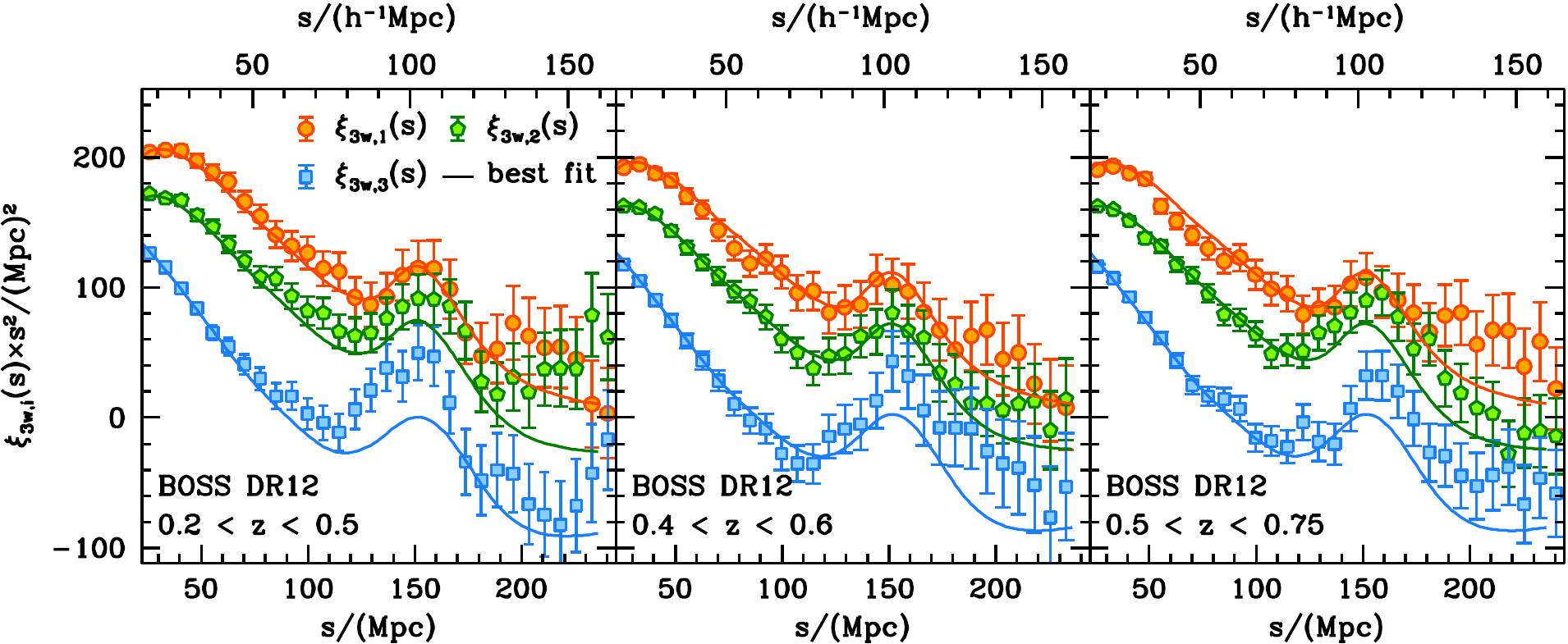}
\caption{
Clustering wedges in the directions parallel (blue) intermediate (green) and transverse (red) to the 
line of sight measured from the combined galaxy sample of BOSS DR12 in our three redshift bins, as a 
function of the pair separation expressed in ${\rm Mpc}$ and $h^{-1}{\rm Mpc}$ in the lower and upper axes, 
respectively. 
The error bars correspond to the dispersion of the results inferred from a set of $N_{\rm m}=2045$  
mock catalogues of the full BOSS survey.
The solid lines correspond to the best-fitting model to these measurements obtained as described in 
Section~\ref{sec:bao}.
}
\label{fig:dr12_3w}
\end{figure*}

Our analysis is part of a series of papers examining the information in the anisotropic clustering
pattern of the combined sample of BOSS DR12. 
\citet{Salazar2016} perform a tomographic analysis of the clustering properties of this sample by 
means of angular correlation functions in thin redshift shells.
\citet{Grieb2016} use the same description of non-linearities, bias and RSD used in our analysis 
to extract cosmological information from the full shape of three clustering wedges measured in Fourier space.
\citet{Satpathy2016} use a model based on convolution Lagrangian perturbation theory 
\citep{Carlson2013,Wang13} and the Gaussian streaming model \citep{Scoccimarro2004, Reid2011} to fit the
full shape of the monopole and quadrupole of the two-point correlation function, $\xi_{0,2}(s)$. 
\citet{Beutler2016b} apply a model based on \citet{Taruya2010} to the power spectrum multipoles
$P_{\ell}(k)$ for $\ell=0,2,4$. 
\citet{Tinker2016} present a comparison of the results of different RSD analysis techniques.
\citet{Ross2016} and \citet{Beutler2016a} perform BAO-only fits to the Legengre multipoles of order 
$\ell=0,2$ of the two-point functions in configuration and Fourier space obtained after the 
application of the reconstruction technique \citep{Eisenstein2007b,Padmanabhan2012} as described
in \citet{Cuesta2016}.
The potential systematics of these BAO-only measurements are discussed in \citet{Vargas2016}.
\citet{Acacia2016} use the methodology described in \citet{Sanchez2016} to combine 
the results presented here with those of the other full-shape and BAO-only analyses into a final set of BOSS 
consensus constraints and explore their cosmological implications.

The outline of this paper is as follows. In Section~\ref{sec:boss} we describe our galaxy sample,
the procedure we follow to measure the clustering wedges, and the mock catalogues used to compute
our estimate of their covariance matrices.
Our model of the full shape of the clustering wedges is described in Section~\ref{sec:model}, 
together with the tests we have performed by applying it to $N$-body simulations and mock catalogues.
In Section~\ref{sec:results} we study the cosmological implications of our clustering measurements.
After describing our methodology to obtain cosmological constraints in Section~\ref{sec:method}, 
Sections~\ref{sec:lcdm} to \ref{sec:gamma} describe the results we obtained
from different combinations of data sets and parameter spaces. 
In Section~\ref{sec:bao} we compress the information of the BOSS clustering wedges into geometric 
constraints and measurements of the growth of structure. Finally, we present our main conclusions in Section~\ref{sec:conclusions}.

\section{The Baryon Oscillation Spectroscopic Survey}
\label{sec:boss}

\subsection{Galaxy clustering measurements from BOSS}
\label{sec:clustering}

We use the final galaxy samples of BOSS, corresponding to SDSS DR12 \citep{Alam2015}. The catalogue is 
divided into two samples, called LOWZ and CMASS, which were selected on the basis of the
SDSS multicolour photometric observations \citep{Gunn1998,Gunn2006}
to cover the redshift range $0.15<z<0.7$ with a roughly uniform comoving number density
$n\simeq3 \times 10^{-4} h^3{\rm Mpc}^{-3}$ \citep{Eisenstein2011,Dawson2013}.
After identifying the galaxies with previous spectroscopic observations from the 
SDSS I/II surveys \citep{York2000}, the remaining redshifts were measured from the spectra obtained with the BOSS spectrographs \citep{Smee2013} as described in \citet{Aihara2011} and \citet{Bolton2012}. 

The CMASS sample is approximately complete down to a limiting stellar mass of
$M\simeq10^{11.3}\,{\rm M}_{\odot}$ for $z>0.45$ \citep{Maraston2013}, with a $\sim$10 per cent satellite
fraction \citep{White2011,Nuza2013}.  
Although it is dominated by early type galaxies, 
$\sim$26 per cent of this sample consist of massive spirals showing star formation activity
in their spectra \citep{Masters2011, Thomas2013}.
The LOWZ sample consists primarily of red galaxies located in massive haloes, and has 
$\sim$12 per cent satellite fraction \citep{Parejko2013}. As described in \citet{Reid2016},
a few regions of the LOWZ sample in
the northern galactic cap (NGC) were targeted using different photometric cuts,
leading to a reduction of the galaxy number density. The obtained galaxy samples, which cover approximately 
1000 deg$^2$, are labelled LOWZE2 and LOWZE3.

Previous clustering analyses of BOSS data have made use of the LOWZ and CMASS samples separately, 
excluding the LOWZE2 and LOWZE3 regions. Here we use the full BOSS data set by combining all these samples
as described in \citet{Reid2016}. 
We follow \citet{Acacia2016} and split this combined sample into three overlapping redshift bins
of roughly equal volume defined by 
$0.2 < z < 0.5$, $0.4 < z < 0.6$ and $0.5 < z < 0.75$.

We study the clustering properties of the combined BOSS galaxy sample by means of 
the clustering wedges statistic \citep{Kazin2012}, $\xi_{\mu_1}^{\mu_2}(s)$,
which corresponds to the average of the full two-dimensional correlation function $\xi(\mu,s)$,
where $\mu$ is the cosine of the angle between the separation vector $\mathbf{s}$ and the line-of-sight
direction, over the interval $\Delta\mu=\mu_{2}-\mu_{1}$, that is
\begin{equation}
\xi_{\mu_1}^{\mu_2}(s)\equiv \frac{1}{\Delta \mu}\int^{\mu_2}_{\mu_1}{\xi(\mu,s)}\,{{\rm d}\mu}.
\label{eq:wedges}
\end{equation}
\citet{Sanchez2013,Sanchez2014} used two wide clustering wedges, dividing the $\mu$ range from 0 to 1 into two
equal-width intervals. Here we measure three wedges, which we denote by $\xi_{3{\rm w}}(s)$ 
and refer to each individual wedge as $\xi_{3{\rm w},i}(s)$ for the intervals $(i-1)/3 < \mu < i/3$. 
In practice, the value of $\mu$ of a given galaxy pair is estimated as the cosine of the angle between the
separation vector, $\mathbf{s}$, and the line-of-sight direction at the midpoint of $\mathbf{s}$.

The observed galaxy redshifts are converted into distances using the same fiducial cosmology 
as in our companion papers, a flat $\Lambda$CDM model with a matter density parameter $\Omega_{\rm m}=0.31$.
This choice is taken into account in our modelling as described in Section \ref{sec:ap_effect}.
We compute the full two-dimensional correlation function $\xi(\mu,s)$ of the combined sample 
in each redshift bin using the estimator of \citet{Landy1993}. We employ a random catalogue
following the same selection function as the combined sample but containing 50 times more objects.
We compute the clustering wedges by averaging the full $\xi(\mu,s)$ over the corresponding $\mu$ intervals.
As in our companion papers, we use a bin size of $ds=5\,h^{-1}{\rm Mpc}$.

 We assign a series of weights to each object in our catalogue.
First, we apply a weight designed to minimize the variance of our
measurements \citep*{Feldman1994} given by 
\begin{equation}
w_{\rm r}({\bf x})=\frac{1}{1+P_{w}\bar{n}({\bf x})},
\label{eq:wradial}
\end{equation}   
where $\bar{n}({\bf x})$ is the expected number density of the catalogue at a
given position ${\bf x}$ and $P_{w}$ is a scale-independent parameter,
which we set to $P_w=10^4\,h^{-3}{\rm Mpc}^3$.
We also include angular weights to account for redshift failures and fibre collisions.
The LOWZE2, LOWZE3 and CMASS samples require additional weights to correct for the systematic
effect introduced by the local stellar density and the seeing of the observations,
as described in detail in \citet{Ross2016}. 

Figure \ref{fig:dr12_3w} shows the resulting wedges $\xi_{3{\rm w}}(s)$ of the DR12 combined sample 
in our three redshift bins as a function of the pair separation expressed in ${\rm Mpc}$ and
${\rm Mpc}/h$ in the lower and upper axes, respectively. 
These measurements and their corresponding covariance matrices (see Section~\ref{sec:covariance}) will 
be made publicly available\footnote{The final URL will appear in the revised version of this paper.}.
The signature of the BAO is clearly visible in all wedges at $s\simeq 150\,{\rm Mpc}$.
The anisotropic clustering pattern generated by redshift-space distortions leads to significant differences
in the amplitude and shape of the three wedges. The solid lines in the same figure correspond to the
best-fitting models obtained as described in Section~\ref{sec:bao}.

\subsection{Covariance matrix estimation}
\label{sec:covariance}

We assume a Gaussian likelihood function for our BOSS clustering measurements given by
\begin{equation}
-2\ln {\cal L}({\bm \xi}|{\bm\theta})=\left({\bm \xi}-{\bm\xi}_{\rm theo}({\bm\theta})\right)^t\bm{\Psi}\left({\bm \xi}-{\bm\xi}_{\rm theo}({\bm\theta})\right)
\end{equation}
where ${\bm \xi}$ is an array containing the measured clustering wedges and 
${\bm\xi}_{\rm theo}({\bm\theta})$
corresponds to our theoretical modelling of these data for the cosmological parameters ${\bm\theta}$.
The evaluation of the likelihood function requires the knowledge of the inverse of the 
covariance matrix, $\bm{\Psi}=\mathbfss{C}^{-1}$,
also known as the precision matrix, which we estimate using the {\sc Multidark-Patchy} ( {\sc MD-Patchy})
BOSS mock galaxy catalogues described in \citet{Kitaura2016}.
These mocks consist of a set of $N_{\rm m}=2045$ independent realizations of the full BOSS survey, corresponding
to the best-fitting $\Lambda$CDM cosmology to the Planck 2013 CMB measurements \citep{Planck2014}.
We computed the wedges $\xi_{3{\rm w}}(s)$ of each mock catalogue in the same way as for the
real BOSS data, and used these measurements to obtain an estimate of the full covariance matrix 
$\hat{\mathbfss{C}}$ of our clustering measurements.
The error bars in Figure~\ref{fig:dr12_3w} correspond to the square root
of the diagonal entries of $\hat{\mathbfss{C}}$.

As a test of the robustness of our results with respect to the details in the estimation of the covariance
matrix we also used an independent set of 1000 Quick Particle Mesh \citep[QPM,][]{White2014} mock 
realizations of the BOSS combined sample. The covariance matrices inferred from the {\sc QPM} and
{\sc MD-Patchy} mocks are consistent and lead to similar results. However, as the {\sc MD-Patchy} mock
samples give a somewhat better match to the clustering properties of the BOSS combined sample than {\sc QPM} \citep{Kitaura2016} and have a significantly larger number of realizations, we based our final
constraints on the covariance matrices inferred from these mock catalogues. 

Our estimates of the covariance matrix are affected by sampling noise due to the finite number of
mock catalogues. Recent studies have provided a clear description of the dependence of the noise
in the estimated covariance matrix on the number of mock catalogues used \citep*{Taylor2013},
its propagation to the derived parameter uncertainties \citep{Dodelson2013,Taylor2014} and
the correct way to include this additional uncertainty in the obtained constraints \citep{Percival2014}.

The first effect that must be taken into account is that when the covariance matrix is 
estimated from a set of independent realizations, the uncertainties in $\hat{\mathbfss{C}}$ and its
inverse follow the Wishart and inverse-Whishart distributions \citep{Wishart1928}, respectively.
As the inverse-Whishart distribution is asymmetric, the inverse of
$\hat{\mathbfss{C}}$ provides a biased estimate of $\bm{\Psi}$.  
This can be corrected for by including a prefactor 
in the estimate of the precision matrix as \citep*{kaufman1967,Hartlap2007}
\begin{equation}
	\hat{\bm{\Psi}}=\left(1-\frac{N_{\rm b}+1}{N_{\rm m}-1}\right)
	\hat{\mathbfss{C}}^{-1},
	\label{eq:d_invcov}
\end{equation}
where $N_{\rm b}$ corresponds to the total number of bins in our measurements.
We restrict our analysis to $20\,h^{-1}{\rm Mpc}< s < 160 \,h^{-1}{\rm Mpc}$
with a bin-width of $ds=5\,h^{-1}{\rm Mpc}$, leading to $N_{\rm b}=84$ for our three
clustering wedges. As our estimates of the 
covariance matrix are based on the $N_{\rm m}=2045$ {\sc MD-Patchy} mock catalogues, 
the factor of equation~(\ref{eq:d_invcov}) is equal to 0.96.
 
Although the estimate of the precision matrix $\hat{\bm{\Psi}}$ of 
equation~(\ref{eq:d_invcov}) is unbiased, it is still affected by noise, 
which should be propagated into the obtained cosmological constraints.
\citet{Percival2014} derived formulae for their impact
on the errors of the cosmological constraints measured by integrating over the likelihood function. 
They demonstrated that, to account for this extra uncertainty, the recovered parameter
constraints must be rescaled by a factor that depends on $N_{\rm b}$, $N_{\rm m}$ 
and the number of parameters included in the analysis, 
$N_{\rm p}$ \citep[see equation~18 in][]{Percival2014}. 
Depending on the parameter space, our choice of range of scales and binning leads to a modest
correction factor of at most 1.6 per cent for the results inferred from the clustering wedges.
The additional uncertainty due to the finite number of mock catalogues could be reduced by
implementing techniques such as covariance tapering \citep{Paz2015} but, as the impact on our
constraints is small, we simply include these correction in our results.

\section{The model}
\label{sec:model_general}

\subsection{Modelling non-linear gravitational evolution, bias and RSD}
\label{sec:model}

The prediction of the clustering wedges for a given cosmology requires 
a model of the full two-dimensional correlation function $\xi(\mu,s)$.
It is convenient to express $\xi(\mu,s)$ as a linear combination of
Legendre polynomials, $L_\ell(\mu)$, as 
\begin{equation}
 \xi(\mu,s)=\sum_{{\rm even}\ \ell} L_{\ell}(\mu)\xi_{\ell}(s),
\label{eq:expansion}
\end{equation}
where the multipoles $\xi_\ell(s)$ are given by
\begin{equation}
\xi_\ell(s)\equiv \frac{2\ell+1}{2}\int^1_{-1} L_\ell(\mu)\xi(\mu,s)\,{\rm d}\mu.
\label{eq:multipoles}
\end{equation}
In order to obtain a description of the multipoles $\xi_\ell(s)$, it is useful to
work with the two-dimensional power spectrum, $P(\mu,k)$. This quantity can also be decomposed in 
terms of Legendre polynomials, with multipoles given by
\begin{equation}
P_\ell(k)\equiv \frac{2\ell+1}{2}\int^1_{-1} L_\ell(\mu)P(\mu,k)\,{\rm d}\mu,
\label{eq:multipoles_pk}
\end{equation}
from which the multipoles $\xi_{\ell}(s)$ can be obtained as
\begin{equation}
\xi_{\ell}(s)\equiv \frac{i^{\ell}}{2\pi^2}\int^{\infty}_{0} P_{\ell}(k) j_{\ell}(ks)\,k^2{\rm d}k,
\label{eq:pl2xil}
\end{equation}
where $j_{\ell}(x)$ is the spherical Bessel function of order $\ell$.  

An accurate model of the full shape of $P(\mu,k)$ must take into account the effects of 
the non-linear evolution of density fluctuations, galaxy bias and RSD. 
We now describe how each of these distortions is taken into account in our model. 

\subsubsection{Non-linear Dynamics}

The accurate modelling of the effects of the non-linear evolution of density fluctuations has been 
the focus of significant work over the last decade or so. In Renormalized perturbation 
theory \citep[RPT;][]{Crocce2006}  and subsequent developments in terms of the multi-point propagator 
expansion~\citep{BerCroSco0811,BerCroSco1206,TarBerNis1211,BerTarNis1211,CroScoBer1212,TarNisBer1301}
the matter power spectrum is written as 
\begin{equation}
P_{\rm NL}(k)=P_{\rm L}(k)\, G(k)^2+P_{\rm MC}(k),
\end{equation}
where the propagator $G(k)$ corresponds to a resummation of all the terms in the perturbation expansion 
that are proportional to the linear spectrum $P_{\rm L}(k)$, and $P_{\rm MC}(k)$ contains mode-coupling contributions (which at $N$ loops involve convolutions over $N$ linear spectra). To an excellent approximation for CDM spectra, the propagator describes the damping of the BAO, while the mode-coupling power describes the shift of the BAO scale ~\citep{CroSco0801,2010ApJ...720.1650S}. Using e.g. the one-loop approximation to the mode-coupling power in this approach has a limited reach in $k$ \citep[see e.g.][]{CroScoBer1212} which is mainly set by the breaking of Galilean invariance \citep{ScoFri9607} due to the fact that the propagator is resummed while the mode-coupling power is not. 
Here we follow the approach of Crocce, Blas \& Scoccimarro (in prep.), who uses Galilean invariance to 
find a resummation of the mode-coupling power consistent with the resummation of the propagator. With 
this approach, dubbed gRPT, it is possible to obtain an improved description down to smaller scales, 
$k \lesssim 0.25\,h\,{\rm Mpc}^{-1}$ for the uncertainties involved in our measurements, see 
Section~\ref{sec:performance} below. 

\subsubsection{Galaxy bias}
\label{sec:bias}

To describe the clustering of galaxies, we write the bias relation between the matter density fluctuations $\delta$ and the 
galaxy density fluctuations, $\delta_{\rm g}$, as in \cite{Chan2012}
\begin{equation}
\begin{split}
\delta_{\rm g} & =  b_1 \delta + {b_2 \over 2} \delta^2 + \gamma_2\, {\cal G}_2  + \gamma_3^- \, \Delta_3{\cal G}
+ \ldots
\end{split}
\label{BiasRel}
\end{equation}
where at cubic order the only term that contributes to the one-loop galaxy power spectrum through the first two multi-point propagators has been written down. The operators  ${\cal G}_2$ and $\Delta_3{\cal G}$ are defined as,
\begin{equation}
{\cal G}_2(\Phi_v) = (\nabla_{ij}\Phi_v)^2-(\nabla^2\Phi_v)^2, 
\label{G2}
\end{equation}
and 
\begin{equation}
\Delta_3{\cal G} =  {\cal G}_2(\Phi)- {\cal G}_2(\Phi_v), \ \ \ \ \ 
\label{D3G}
\end{equation}
where $\Phi$ and $\Phi_v$ are the normalized density and velocity potentials $\nabla^2 \Phi = \delta$ and   $\nabla^2 \Phi_v = \theta$. 

A few points about the bias relation in Eq.~(\ref{BiasRel}) are worth making here. First, under local-Lagrangian bias, the non-local bias parameters are related to the linear bias $b_1$ as \citep{Fry9604,CatLucMat9807,CatPorKam0011,Chan2012}
\begin{align}
\gamma_2 &=-{2\over 7} (b_1-1), \label{eq:local_lag} \\ 
\gamma_3^- &={11 \over 42} (b_1-1), 
\label{gammasLL}
\end{align}
Second, while there is no compelling argument for the validity of local Lagrangian bias~\citep{SheChaSco1304}, a bispectrum analysis of dark matter halos shows that the  $\gamma_2(b_1)$ relation in Eq.~(\ref{eq:local_lag}) is at least a reasonable first approximation~\citep{Chan2012,BalSelDes1201,SheChaSco1304,SaiBalVla1405,BelHofGaz1504}. In our context here this is particularly relevant given that in this work we use two-point statistics alone, which do not constrain $\gamma_2$ that well. Therefore, we assume the  $\gamma_2(b_1)$ relation in Eq.~(\ref{eq:local_lag}). We have in fact checked using CMASS-type galaxies in the Minerva simulations discussed below by relaxing this assumption that it does not bias our results. 
 
Finally, the situation is somewhat different for the $\gamma_3^-$ parameter, and we {\em do not} assume the $\gamma_3^-(b_1)$ relation in Eq.~(\ref{gammasLL}) for a number of reasons. First, the linear bias $b_1$ is the only bias parameter that receives significant signal to noise over a broad range of scales, as opposed to the rest of the terms in Eq.~(\ref{BiasRel}) that only enter through loop corrections for our two-point function only analysis. Therefore one should in principle include the running of $b_1$ with scale which corresponds to adding a $\nabla^2 \delta$ term in Eq.~(\ref{BiasRel}). However, such term is fairly degenerate with the contribution coming from $\gamma_3^-$ \citep{McDRoy0908,SaiBalVla1405,BiaDesKeh1405} and thus provided we let $\gamma_3^-$ (and $b_2$ as well) be free one can absorbe such contributions given the range of scales considered in our analysis. The same holds for stress tensor contributions to dark matter clustering~\citep{PueSco0908,PieManSav1108,CarHerSen1206,BauNicSen1207} that are fully degenerate with the running of the linear bias. 

Summarizing, our bias model has three free parameters corresponding to $b_1,b_2,\gamma_3^-$, with $\gamma_2$ given in terms of $b_1$ by the local Lagrangian bias relation in Eq.~(\ref{eq:local_lag}). For detailed expressions of the galaxy power spectrum that follow from the equations above, see Appendix~\ref{app:details}.

\subsubsection{Redshift-space distortions}
\label{sec:rsd}

We base our description of the redshift-space power spectrum on~\citep{Scoccimarro1999}
\begin{equation}
\begin{split}
P(k,\mu) &= \int {d^3r \over (2\pi)^3} {\rm e}^{-i \kk \cdot \r}\, W(\lambda,\r) \Big[ \langle {\rm e}^{\lambda \Delta  u_z} D_s D_s' \rangle_c \\
        &\quad  
          +  \langle {\rm e}^{\lambda \Delta  u_z} D_s \rangle_c + \langle {\rm e}^{\lambda \Delta  u_z} D_s' \rangle_c
\Big],
\end{split}
\label{SCF99}
\end{equation}
where $\lambda = i f k \mu$, $W(\lambda,\r)= \langle {\rm e}^{\lambda \Delta  u_z} \rangle_c$ is the generating function of velocity differences, and $D_s \equiv \delta_g + f \nabla_z u_z$, with a prime denoting a quantity
at $\x'$ instead of $\x$, and $\r=\x-\x'$. 
In the Gaussian approximation, the generating function can be written as
\begin{equation}
W_G(\lambda,\r)={\rm e}^{\lambda^2(\sigma_v^2 - \psi_\perp+\nu^2 \Delta\psi)},
\label{Wgau}
\end{equation}
where $\psi_\perp = (I_0+I_2)/3$, $\Delta\psi = I_2$, $\sigma_v^2=\psi_\perp(0)$ and 
\begin{equation}
 I_\ell(r) \equiv \int d^3k \, j_\ell(kr)\, \frac{P(k)}{k^2}.
\label{Iell}
\end{equation}
In the large-scale limit  $W_G(\lambda,\r\to\infty)={\rm e}^{\lambda^2 \sigma_v^2}$ becomes scale-independent. However, as pointed out in ~\cite{Scoccimarro2004} it is necessary to include non-linear corrections to this factor,
which correspond mostly to fingers-of-God (FOG) or virial motions, since the large-scale limit of the velocity distribution function {\em is not} Gaussian. Therefore instead of $W_G(\lambda,\r\to\infty)$ we use,
\begin{equation}
W_\infty(\lambda) = {1  \over \sqrt{1-\lambda^2 a_{\rm vir}^2}}\, \exp \Big({ \lambda^2 \sigma_v^2 \over {1-\lambda^2 a_{\rm vir}^2}} \Big),
\label{W0}
\end{equation}
where $a_{\rm vir}$ is a free parameter that describes the contribution of small-scale velocities and
characterizes the kurtosis of the velocity distribution, while $\sigma_v$ is predicted as above. 
This is thus the form of our FOG factor, which can be obtained by resumming quadratic nonlinearities as advocated in ~\cite{Scoccimarro2004}.
To calculate
the expression in square brackets (whose Fourier transform corresponds roughly to a ``no-virial" power spectrum)
we use the one-loop approximation,
\begin{equation}
\begin{split}
P_{\rm novir}(k,\mu) &= \int {d^3r \over (2\pi)^3} {\rm e}^{-i \kk \cdot \r}\,  \Big[ \langle D_s D_s' \rangle_c
+ \lambda  \langle  \Delta  u_z D_s D_s' \rangle_c \\
&\quad + \lambda^2 \langle \Delta  u_z D_s \rangle_c \langle \Delta  u_z D_s' \rangle_c \Big].
\end{split}
\label{Pnovir}
\end{equation}
Therefore, the result for the redshift-space power spectrum is given by
\begin{equation}
P(k,\mu) = W_\infty(if k \mu) \, P_{\rm novir}(k,\mu),
\label{Prsd}
\end{equation}
and multipoles can be obtained directly by integrating this equation against Legendre polynomials
$L_\ell(\mu)$ as in equation~(\ref{eq:multipoles_pk}).
We now briefly describe how we calculate each of the terms in equation~(\ref{Pnovir}). 
A more detailed description of the involved terms can be found in Appendix~\ref{app:details}.

The first term involving $\langle D_s D_s' \rangle_c$ is simply the non-linear version of the well-known
Kaiser formula \citep{Kaiser1987},
\begin{equation}
P_{\rm novir}^{(1)}(k,\mu) = P_{gg}(k) + 2 f \mu^2 P_{g\theta}(k) + f^2 \mu^4 P_{\theta\theta}(k)
\label{Pkaiser}
\end{equation}

Assuming that there is no velocity bias, $P_{\theta\theta}(k)$ can be obtained directly from the predictions
of gRPT. Appendix~\ref{app:details} contains explicit formulae for $P_{gg}(k)$ and $P_{g\theta}(k)$.

The term involving $\langle  \Delta  u_z D_s D_s' \rangle_c$ in equation~(\ref{Pnovir}) is to leading order given by the tree-level bispectrum between densities and velocities as
\begin{equation}
\begin{split}
P_{\rm novir}^{(2)}(k,\mu) &= \int {q_z\over q^2} \Big[ B_{\theta D_s D_s}(\q,\kk-\q,-\kk) \\
                           &\quad + B_{\theta D_s D_s}(\q,-\kk,\kk-\q)\Big],
\label{Pnovir2}
\end{split}
\end{equation}
with the bispectra given by standard tree-level PT for densities and velocities in terms of the 
$F_2$ and $G_2$ kernels and bias parameters $b_1,b_2, \gamma_2$.

The term $\langle \Delta  u_z D_s \rangle_c \langle \Delta  u_z D_s' \rangle_c$ in equation~(\ref{Pnovir}) is already quadratic in the power spectrum, so this can be evaluated using linear perturbation theory. 
We then have,
\begin{equation}
\begin{split}
P_{\rm novir}^{(3)}(k,\mu) &= \int {q_z\over q^2} {(k_z-q_z)\over (\kk-\q)^2} (b_1+f \mu_q^2)(b_1+f \mu_{k-q}^2) \\
                           &\quad \times P_{\delta\theta}(k-q)P_{\delta\theta}(q) d^3q.
\end{split}
\label{Pnovir3}
\end{equation}

The closest redshift-space model in the literature to ours~\citep{Taruya2010,BeuSaiSeo1409} also 
starts from equation~(\ref{SCF99}). Our approach has three main differences, namely, we include 
nonlinear bias contributions coming from $b_2$ and $\gamma_2$ to the bispectra in Eq.~(\ref{Pnovir2}), 
our FOG factor Eq.~(\ref{W0}) is non-Gaussian, we let $\gamma_3^-$ be a free parameter (instead of 
being fixed to its local Lagrangian bias value), and we use gRPT to calculate matter loops 
instead of RegPT (which is not Galilean invariant). 
In summary, note that our redshift-space model has a single free parameter, $a_{\rm vir}$. 
It can be considered as the large-scale limit to a more complete model in which the velocity 
dispersion is scale-dependent and other small-scale effects are included (Scoccimarro in prep.). 
The main reason for these simplifications is that the model as presented here can be numerically 
evaluated very efficiently for cosmological parameter estimation.

\subsection{The Alcock-Paczynski effect}
\label{sec:ap_effect}

\begin{figure}
\includegraphics[width=0.45\textwidth]{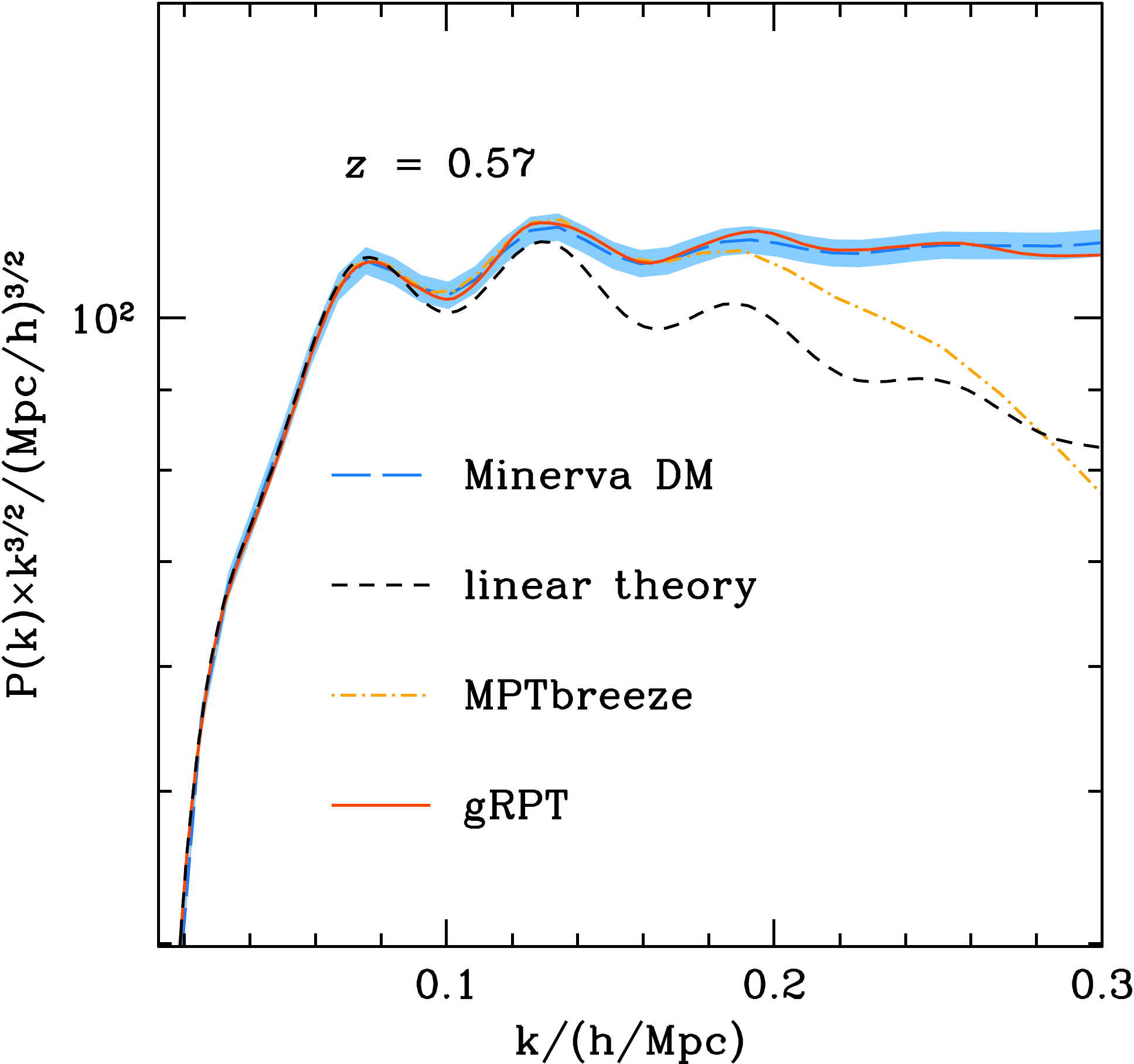}
\caption{
Mean dark-matter real-space power spectrum of the Minerva simulations at $z = 0.57$ (blue 
long-dashed lines) compared against the predictions of linear theory (black short-dashed lines), 
two-loop RPT as implemented in {\sc MPTBreeze} (orange dot-dashed lines) and 
one-loop gRPT (red solid lines). The shaded region corresponds to a 2 per cent uncertainty in 
the value of $P(k)$.
}
\label{fig:minerva_pk}
\end{figure}

As described in Section~\ref{sec:boss}, clustering measurements from real galaxy catalogues depend
on the assumption of a fiducial cosmology used to transform the observed redshifts into comoving
distances. 
Assuming a fiducial cosmology that deviates from the true underlying one leads to a
rescaling of the components parallel and perpendicular to the line-of-sight, 
$s_{\parallel}$ and $s_{\perp}$, of the total separation vector ${\mathbf s}$ between two galaxies as 
\citep{Padmanabhan2008,Kazin2012}
\begin{align}
 s_{\perp}     &= q_{\perp}s'_{\perp},\label{eq:scaling1}\\
 s_{\parallel} &= q_{\parallel}s'_{\parallel},\label{eq:scaling2}
\end{align}
where the primes denote the quantities in the fiducial cosmology and the scaling factors are given by
the ratios of the angular diameter distance and the Hubble parameter in the true and fiducial
cosmologies at the mean redshift of the sample, $z_{\rm m}$, as
\begin{align}
 q_{\perp} &= \frac{D_{\rm M}(z_{\rm m})}{D'_{\rm M}(z_{\rm m})}, \label{eq:q_perp}\\
 q_{\parallel} &= \frac{H'(z_{\rm m})}{H(z_{\rm m})},\label{eq:q_para}
\end{align}
Equations (\ref{eq:scaling1}) and (\ref{eq:scaling2}) are the basis of the the Alcock--Paczynski test 
\citep{Alcock1979}, which allows for anisotropic BAO measurements \citep{Hu2003,Blake2003,Linder2003}.
In terms of $s$ and $\mu$, these equations can be written as \citep{Ballinger1996}
\begin{align}
 s &= s'q(\mu'),\label{eq:sfid}\\
\mu &= \mu'\frac{q_{\parallel}}{q(\mu')},\label{eq:mufid}
\end{align}
where
\begin{equation}
q(\mu)=\left[q_{\parallel}^2(\mu')^2+q_{\perp}^2(1-(\mu')^2)\right]^{1/2}. \label{eq:alpha_of_mu}
\end{equation}
The scaling factors of equations (\ref{eq:q_perp}) and (\ref{eq:q_para}) are often denoted 
$\alpha_{\perp,\parallel}$.  
However, we will reserve that notation for the combination of these purely geometric quantities
with the sound horizon ratios in the fiducial and true cosmology, as described in \citet{Acacia2016}.
For historical reasons, most clustering measurements are expressed in units of $h^{-1}{\rm Mpc}$. As the
value of $h$ of a given cosmological model will in general be different from that of the fiducial
cosmology, the ratios of equations (\ref{eq:q_perp}) and (\ref{eq:q_para}) must also be computed in these units. 

\begin{figure}
\includegraphics[width=0.45\textwidth]{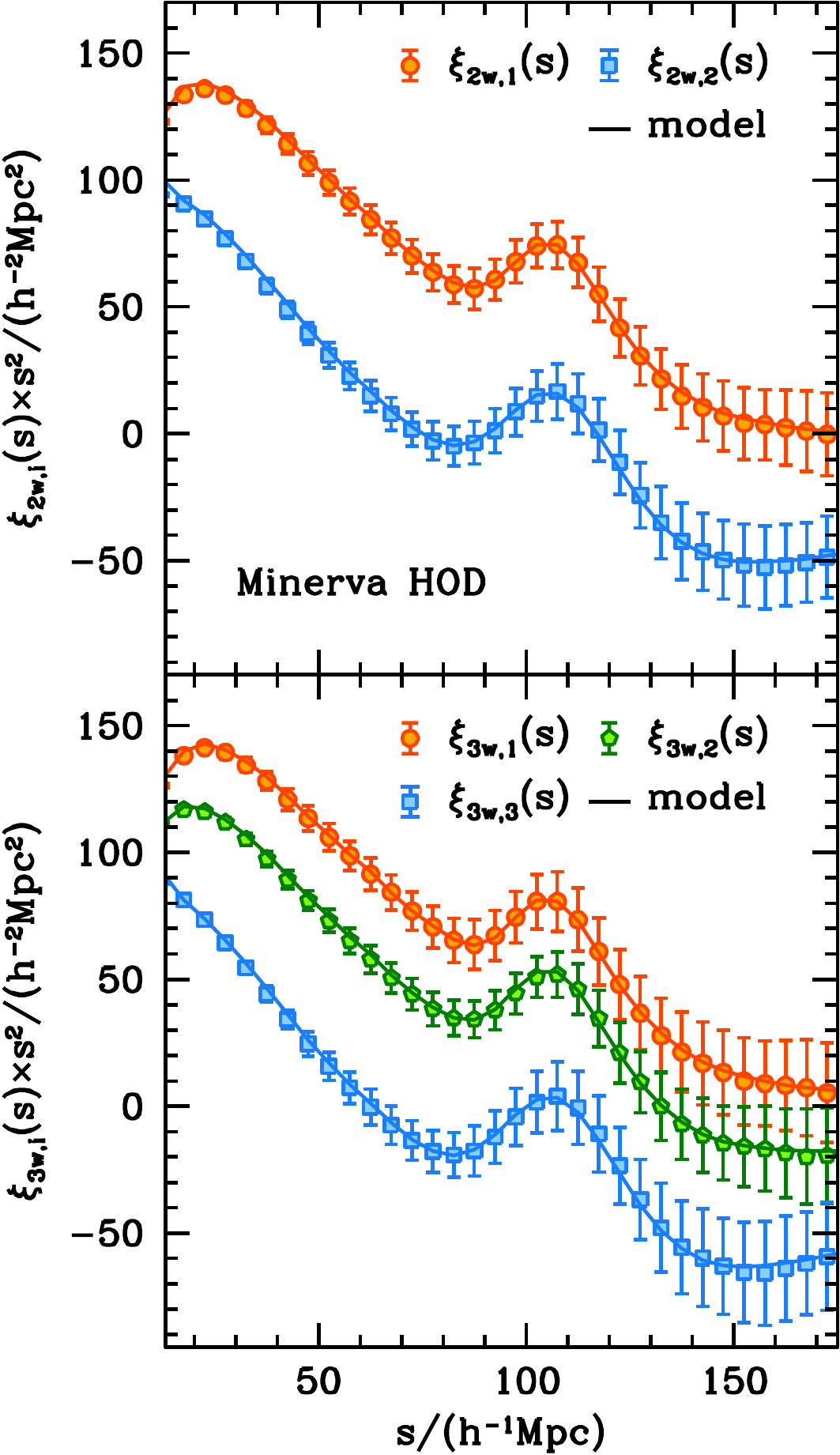}
\caption{
Mean clustering wedges of the {\sc Minerva} HOD samples for the two (upper panel) and three (lower
panel) $\mu$-bins configurations. The error bars correspond to the square root of the
diagonal entries of the covariance matrices computed using the Gaussian recipes of \citet{Grieb2016}.
The solid lines correspond to the model described in Section \ref{sec:model}, which gives an excellent
description of the simulation results.
}
\label{fig:minerva}
\end{figure}

Before comparing the predictions of a given cosmological model with our BOSS clustering measurements we use equations~(\ref{eq:sfid}) and (\ref{eq:mufid}) to transform our model of $\xi(\mu,s)$ to the fiducial cosmology assumed in their estimation by expressing the integral in equation (\ref{eq:wedges}) as
\begin{equation}
\xi_{\mu_1}^{\prime\mu_2}(s')\equiv \frac{1}{\mu'_2-\mu'_1}\int^{\mu'_{2}}_{\mu'_{1}}\xi(\mu(\mu',s'),s(\mu',s'))\,{{\rm d}\mu'}.
\label{eq:wedges_fid}
\end{equation}

\subsection{Performance of the model}
\label{sec:performance}

\subsubsection{Minerva simulations}

\begin{figure}
\includegraphics[width=0.45\textwidth]{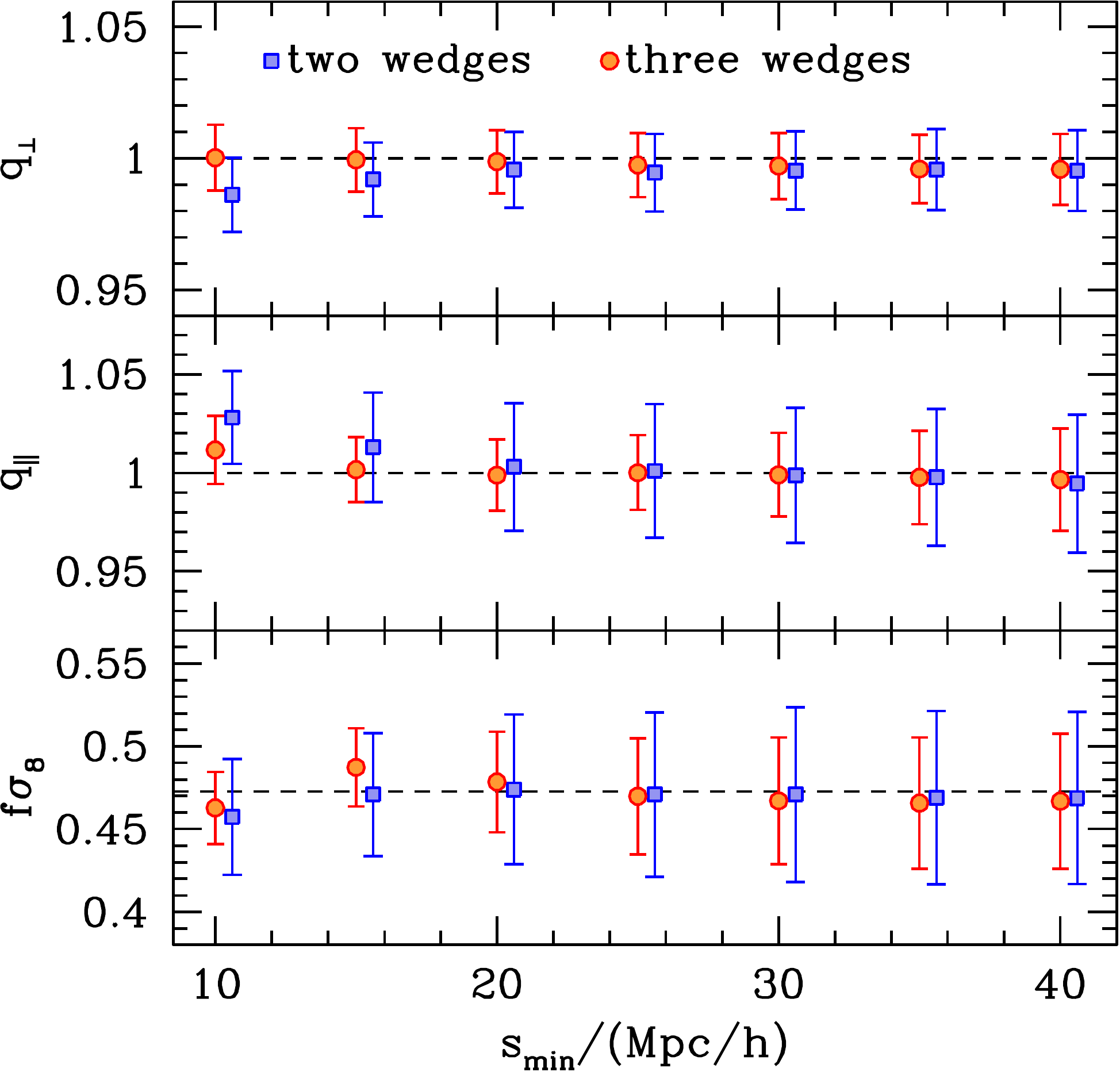}
\caption{
Mean values (points) and 68 per cent CL on $q_{\perp}$, $q_{\parallel}$ and $f\sigma_8$ derived
from the measurements of two (squares) and three (circles) clustering wedges from the {\sc Minerva} HOD
galaxy samples as a function of the minimum scale included in the fits. The dashed lines correspond to
the true values of these parameters.  
Based on this test we set a minimum scale of $s_{\rm min}=20\,h^{-1}{\rm Mpc}$ for our fits to the BOSS
combined sample clustering wedges.
}
\label{fig:alpha_smin}
\end{figure}

To evaluate the performance of the model described in Sec. \ref{sec:model} we
used a set of 100 N-body simulations called {\sc Minerva}, which are described in more 
detail in \citet{Grieb2016}.  
These simulations represent different realizations of the same 
cosmology, corresponding to the best-fitting flat $\Lambda$CDM model to the 
combination of CMB data and the wedges of the CMASS sample from SDSS DR9 from \citet{Sanchez2013}. 
This model is characterized by a matter density of $\Omega_{\rm m}= 0.285$, a baryon physical
density of $\omega_{\rm b}=0.02224$, a Hubble constant of 
$H_0=69.5\,{\rm km}{\rm s}^{-1}{\rm Mpc}^{-1}$, a scalar spectral
index of $n_{\rm s} = 0.968$ and an amplitude of density fluctuations of $\sigma_8 = 0.828$.
Each simulation traces the evolution of the
dark-matter density field with $N_{\rm part}=1000^3$ over a box of side length 
$L_{\rm box}=1.5\,{\rm Gpc}/h$. 
The initial conditions were generated with second-order Lagrangian perturbation theory (2LPT)
at a starting redshift of $z_{\rm ini}=63$.

 Figure~\ref{fig:minerva_pk} shows a comparison of the mean dark-matter real-space power spectrum of the
{\sc Minerva} simulations at $z=0.57$ with the predictions of RPT (dashed lines) computed
using {\sc MPTBreeze} \citep{Crocce2012},
and one-loop gRPT (solid lines). The shaded regions corresponds to a 2\% uncertainty in the value
of $P(k)$. The prediction from RPT is in good agreement with the simulation results up to
$k \lesssim 0.15\,h\,{\rm Mpc}^{-1}$, and describe accurately the damping of the first BAO peaks.
Using gRPT, the description of the simulation results can be extended up to modes as high as
$k\lesssim 0.25\,h\,{\rm Mpc}^{-1}$. As our model of the full shape of the clustering wedges 
is based on gRPT, we can expect to be able to extend the range of scales included in our analysis
with respect to the analyses of \citet{Sanchez2013,Sanchez2014}.

In order to extend these models to real galaxy clustering measurements it is necessary to include the 
effects of bias and RSD.
We model galaxy and halo bias including both local and non-local contributions given by the parameters
$b_1$, $b_2$, $\gamma_2$ and $\gamma_3^{-}$ defined in Section~\ref{sec:bias}. 
As our two-point clustering measurements are not significantly sensitive to $\gamma_2$ we use the
local-Lagrangian relation of equation (\ref{eq:local_lag}) to set its value in terms of $b_1$ and treat the remaining quantities as free parameters.  

\begin{figure}
\includegraphics[width=0.45\textwidth]{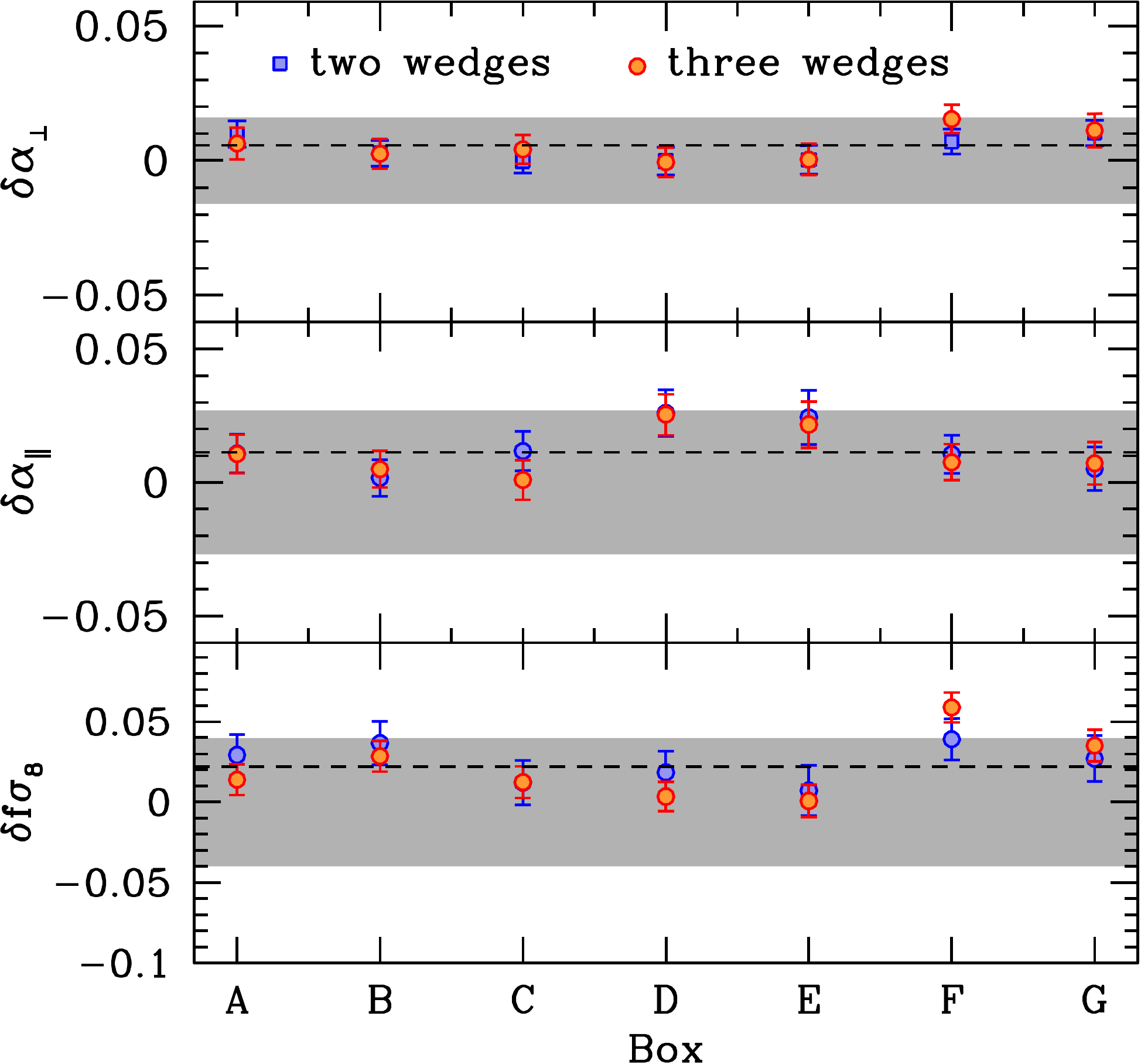}
\caption{
Difference between the values of $\alpha_{\perp}$, $\alpha_{\parallel}$ and $f\sigma_8$ 
obtained from the measurements of two (squares) and three (circles) wedges from each of the 
HOD boxes (labelled A to G) of the RSD challenge of \citet{Tinker2016}. The dashed lines correspond
to the mean differences over all boxes. The shaded regions indicate the uncertainties associated
with the constraints on these parameters inferred from the real BOSS sample (see Section~\ref{sec:bao}).
}
\label{fig:chal_boxes}
\end{figure}

We used the snapshots at $z=0.57$ of the {\sc Minerva} 
simulations, corresponding to the mean redshift of the CMASS sample, in which we identified bound 
halos using a friends-of-friends algorithm. 
The resulting sample was later post-processed with {\sc Subfind} \citep{Springel2000} to eliminate 
spurious unbound objects, leading to a final halo catalogue with a minimum mass of 
$M_\mathrm{min} = 2.67 \times 10^{12} \; h^{-1} \,{\rm M}_{\odot}$.  
\citet{Grieb2016} populated the Minerva halo catalogues at $z=0.57$ with 
galaxies following a halo occupation distribution (HOD) model parametrized as in \citet{Zheng2007},
in order to match the monopole correlation function of the CMASS sample. 
The values of the parameters characterizing this HOD are similar to those used by \citet{Manera2013},
but the mass resolution of the {\sc Minerva} simulations allows us to resolve the halos of the low-mass
tail of the distribution. The clustering properties of the resulting HOD galaxy samples closely match
those of the CMASS sample of BOSS. 
We use these HOD catalogues to test if our full model of equation (\ref{Prsd}) correctly describes 
the effect of non-linear evolution, bias and RSD, including the impact of the FOG effect, on a 
sample that contains both central and satellite galaxies.

The points in Fig.~\ref{fig:minerva} correspond to the mean wedges from the HOD galaxies of
all {\sc Minerva} realizations for two (upper panel) and three $\mu$-bins (lower panel) configurations.
As the 100 {\sc Minerva} realizations are not enough to obtain a robust estimate of the covariance matrix
of these measurements, we use the Gaussian recipes of \citet{Grieb2016}, computed using the
multipoles of the non-linear power spectrum model of Section \ref{sec:model} as input. 
The error bars in Fig.~\ref{fig:minerva} correspond to the square root of the
diagonal entries of the resulting covariance matrices.
As shown by \citet{Grieb2016}, these Gaussian formulae give an excellent description of the results
inferred from the {\sc Minerva} simulations.  
Using these covariance matrices, we fitted for the nuisance parameters of the model using the 
measurements of two clustering wedges, while 
fixing all cosmological parameters to their true values. The solid lines in Fig.~\ref{fig:minerva} 
correspond to the model described in Section \ref{sec:model}, computed using the resulting values
for the nuisance parameters, which show an excellent agreement with the
results from the {\sc Minerva} simulations up to small scales.

\begin{figure}
\includegraphics[width=0.47\textwidth]{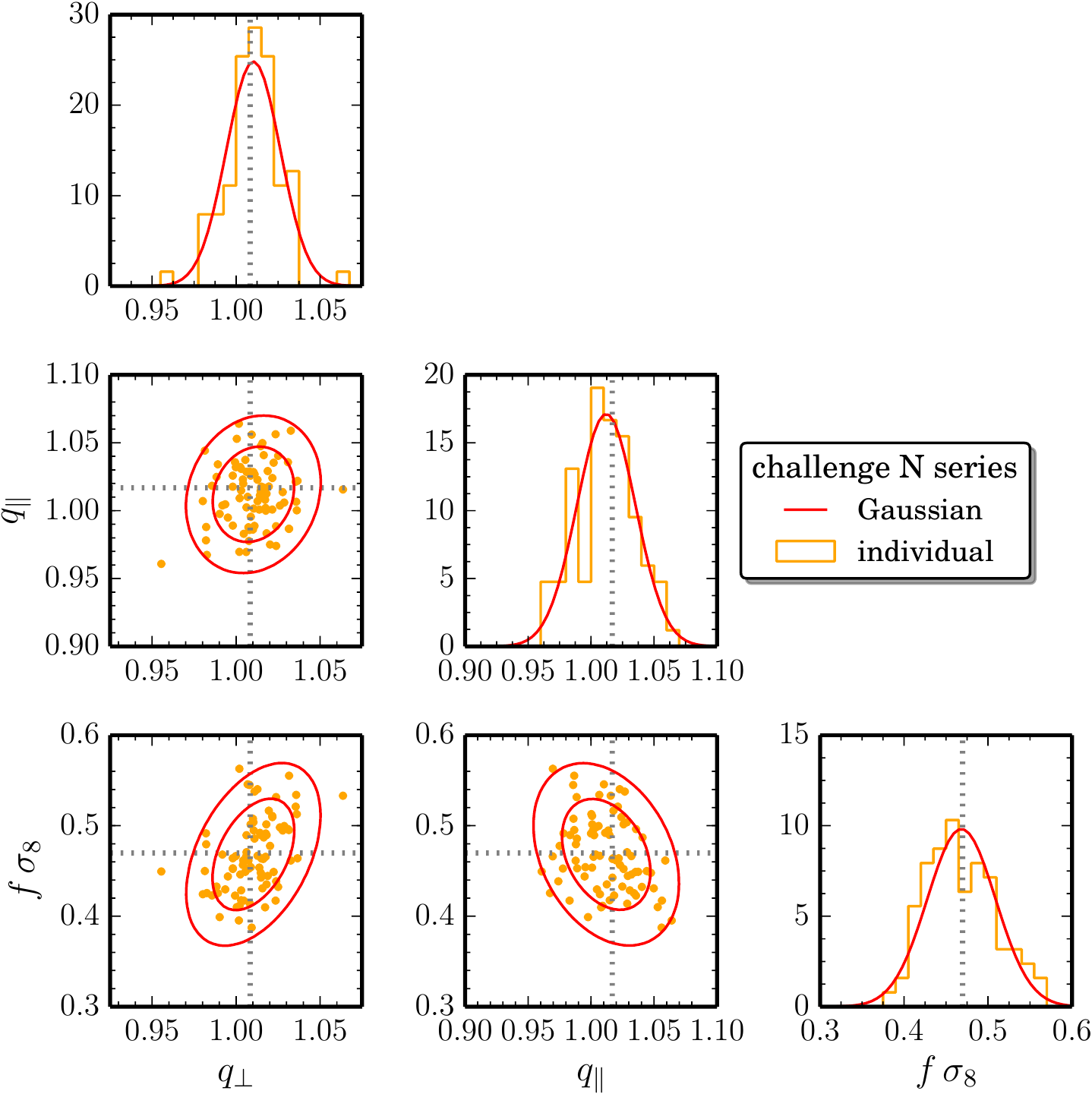}
\caption{
Constraints on $q_{\perp}$, $q_{\parallel}$ and $f\sigma_8$ obtained from the 83 CMASS mock 
catalogues of the RSD challenge of \citet{Tinker2016}. 
The points in the off-diagonal panels correspond to the values recovered from the individual mocks, 
while the histograms in the diagonal panels show the distribution of the obtained results from the 
full set of mocks. The red solid lines correspond to a Gaussian fit to the obtained distribution.
}
\label{fig:cutsky}
\end{figure}
 
In order to test the ability of our model to provide unbiased cosmological constraints 
we treated the quantities $q_{\perp}$, $q_{\parallel}$ and $f\sigma_8$
as free parameters and fit for them using the mean clustering wedges from the {\sc Minerva}
simulations, varying simultaneously the nuisance parameters of the model while fixing all
cosmological parameters
to their correct values (i.e. fixing the shape of the linear-theory power spectrum).
Fig.~\ref{fig:alpha_smin} shows the obtained constraints for the cases of two (squares) and three (circles) wedges as a function of the minimum scale included in the fits, $s_{\rm min}$. 
The points indicate the mean values of these parameters derived from our MCMC while the error
bars correspond to their respective 68\% confidence levels (CL). In all cases the maximum scale was set to 
$s_{\rm max}=160\,h^{-1}{\rm Mpc}$. 
The dashed lines in the same Figure correspond to the true values of these parameters.  

The constraints obtained using both configurations are in perfect agreement with the
true underlying values of these parameters, but the 68\% C.L. obtained with three clustering wedges are significantly smaller than those recovered from the analysis of two $\mu$-bins. This clearly 
illustrates the power of the additional information recovered from 
three clustering wedges, with respect to that of using only two.
These results are consistent with those of the Fourier-space analysis of \citet{Grieb2016}, which is based on the same underlying model of non-linearities, bias and RSD.
As $s_{\rm min}$ is reduced, the allowed ranges for all parameters decrease. The results from this test indicate that the application of the model described in Section~\ref{sec:model} to a measurement of three clustering wedges can give unbiased cosmological constraints even when including scales as small as
$s_{\rm min} \simeq 15\,h^{-1}{\rm Mpc}$. As this limit might depend on the details of the cosmological model, we fixed the value of $s_{\rm min} = 20 \,h^{-1}{\rm Mpc}$ for our analysis of the clustering wedges from the BOSS combined galaxy sample.  

\begin{figure}
\includegraphics[width=0.45\textwidth]{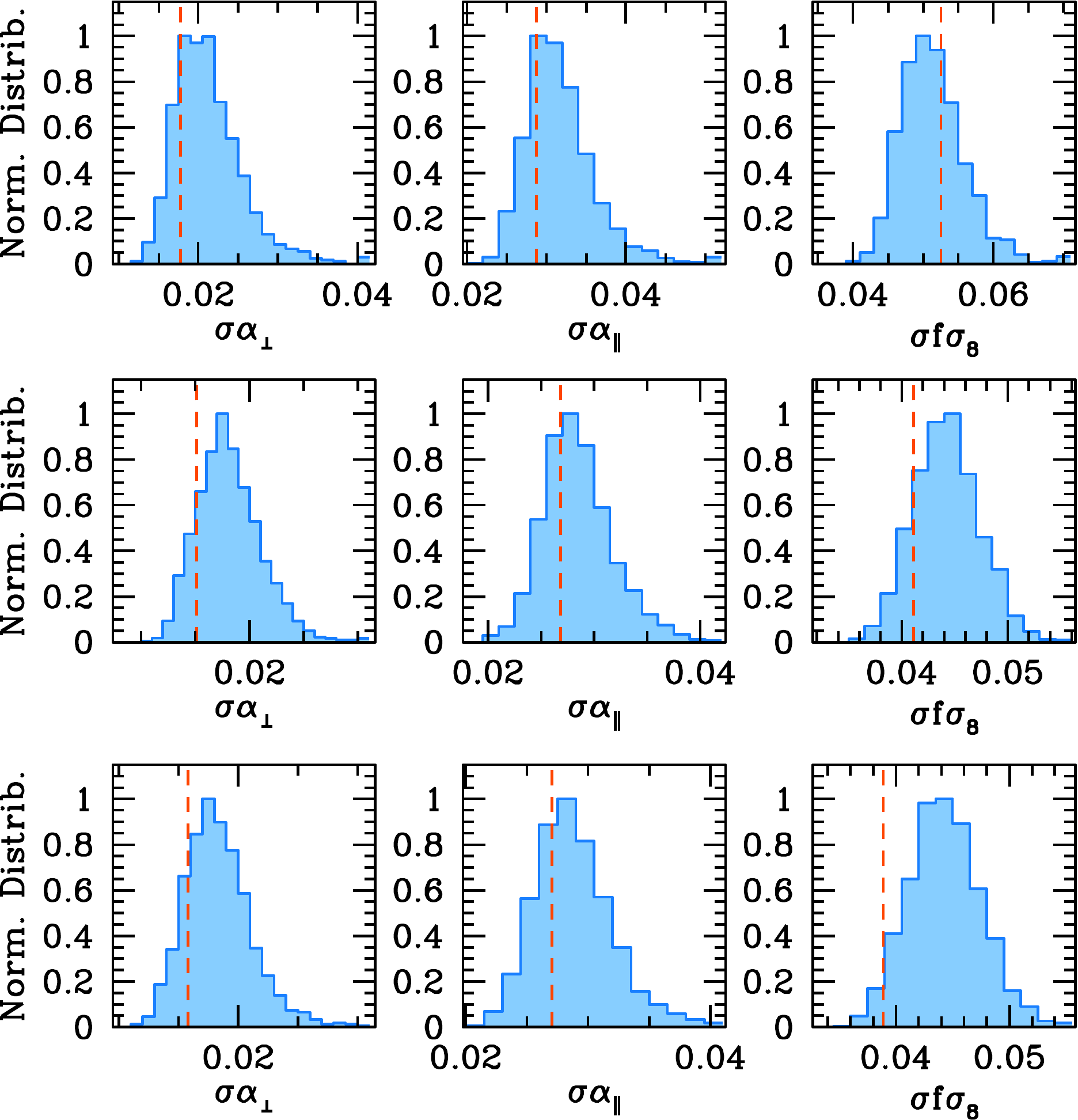}
\caption{
Distributions of the marginalized 68\% CL on the values of the parameters $\alpha_{\perp}$, 
$\alpha_{\parallel}$ and $f\sigma_8$ obtained from the individual {\sc MD-Patchy} mocks in each of our 
three redshift bins. The vertical dashed lines indicate uncertainties on these parameters obtained 
from the real BOSS clustering wedges (see Section~\ref{sec:bao}).
}
\label{fig:histo_patchy}
\end{figure}

\subsubsection{The BOSS RSD challenge}

Our companion paper \citet{Tinker2016} presents the results of a comparison
or ``challenge'' of various RSD models and methodologies to extract cosmological 
information from the full shape of anisotropic clustering measurements. 
This challenge consisted of two different tests: an ensemble of 83 mock
catalogues of the NGC CMASS sub-sample, and a series of seven
simulation boxes corresponding to different cosmologies and HOD parametrizations.
A more detailed description of these data sets and the results obtained by the
different methods can be found in \citet{Tinker2016}. Here we summarize the 
results obtained by applying the model described in Section~\ref{sec:model} to the
measurements of three clustering wedges in configuration space obtained from these
data sets. 

Figure~\ref{fig:chal_boxes} shows the difference between the values of 
$\alpha_{\perp}$, $\alpha_{\parallel}$ and $f\sigma_8$ recovered from the measurements 
of $\xi_{3{\rm w}}(s)$ from each of the seven HOD boxes, labelled A to G 
\citep[see ][for details on the HOD applied in this case]{Tinker2016}.
The dashed lines correspond to the mean differences over all boxes.
A covariance matrix derived from a set of 1000 quick particle mesh \citep[QPM,][]{White2014} 
simulations with a box size of 
$2.5\,h^{-1}{\rm Gpc}$ and an HOD matching the clustering of the CMASS sample was
used for all the fits. As these results correspond to different
cosmologies and HODs, it is not possible to derive a general conclusion about the expected
deviation between the true and obtained results. However, with the exception of the value of
$f\sigma_8$ recovered from box F, the obtained deviations are always smaller than the
uncertainties with which these parameters can be recovered from
the BOSS sample (see Section~\ref{sec:bao}), which are indicated by the grey shaded regions.

Figure~\ref{fig:cutsky} summarizes the constraints on $q_{\perp}$, $q_{\parallel}$ and 
$f\sigma_8$ obtained from the set of 83 CMASS mock catalogues. 
The points in the off-diagonal panels correspond to the recovered values of these parameters 
from each individual realizations, while the histograms in the diagonal panels show the distribution
of the obtained results from the full set of mocks. 
The red solid lines correspond to the Gaussian fit to the obtained distribution.
The constraints obtained using our methodology are in excellent agreement with
the true underlying values of these parameters indicated by the dotted lines.

\subsubsection{The {\sc MD-Patchy} mock catalogues}
\label{sec:patchy_rsd}

\begin{table} 
\centering
  \caption{Mean and dispersion of the deviations between the parameter constraints obtained from the individual
{\sc MD-Patchy} mock catalogues and their true underlying values for our three redshift bins.}
\label{tab:results_patchy}
\begin{tabular}{cccc}
\hline
  Parameter                & $0.2 < z < 0.5$  & $0.4 < z < 0.6$  & $0.5 < z < 0.75$    \\
\hline
$\delta\alpha_{\perp}$     & $0.003\pm0.022$  & $0.001\pm0.018$ & $0.001\pm0.018$       \\
$\delta\alpha_{\parallel}$ & $0.006\pm0.032$  & $0.005\pm0.027$ & $0.005\pm0.028$       \\
$\delta f\sigma_8$         & $-0.018\pm0.052$ & $0.009\pm0.044$ & $0.004\pm0.044$        \\
\hline
\end{tabular}
\label{tab:results_patchy}
\end{table}

As a final test of our model we applied to the measurements of $\xi_{3{\rm w}}(s)$
from each of the 2045 {\sc MD-Patchy} mocks of the BOSS DR12 combined sample described in 
Section~\ref{sec:covariance}. Besides providing another test for possible systematic
errors in our constraints, the obtained values can give us an idea of the uncertainties 
we can expect to obtain from the analysis of the real BOSS data.
These constraints are also used in \citet{Sanchez2016} to compute the cross-correlation
coefficients between the results inferred from $\xi_{3{\rm w}}(s)$ and those of our
companion papers. 

Table~\ref{tab:results_patchy} lists the mean and dispersion of the difference between 
values of $\alpha_{\perp}$, $\alpha_{\parallel}$ and $f\sigma_8$ obtained from the {\sc MD-Patchy}
mocks and their correct values in each of our three redshift bins. Deviations of the order of 
0.3$\sigma$ and 0.2 $\sigma$can be seen in the value of $f\sigma_8$ obtained using data from the low- 
and intermediate-redshift bins, respectively. 
Although this might indicate the presence of a small systematic error in 
these measurements, as these differences are much smaller than their associated statistical errors we 
do not include a systematic uncertainty in our results. 

Figure~\ref{fig:histo_patchy} shows the distributions of the marginalized 68\% C.L.
on the values of $\alpha_{\perp}$, $\alpha_{\parallel}$ and 
$f\sigma_8$ obtained from the Patchy mocks in the low- (upper panels), intermediate- (middle panels)
and high-redshift (lower panels) bins.
The vertical dashed lines indicate the uncertainties on these parameters obtained from the 
real BOSS clustering wedges as described in Section~\ref{sec:bao}, which are in good agreement
with the distributions obtained from the {\sc MD-Patchy} mocks.

\section{Cosmological implications}
\label{sec:results}

\subsection{Methodology for parameter constraints}
\label{sec:method}

\begin{table} 
  \caption{
Cosmological parameters constrained in our analysis.
The upper part lists the parameters of the standard $\Lambda$CDM model while the middle section 
lists a number of its possible extensions. The lower part list a number of additional quantities
whose values can be derived from the first two sets.
}
    \begin{tabular}{ll}
    \hline
 \multicolumn{1}{c}{Parameter} & \multicolumn{1}{c}{Description} \\
\hline
\multicolumn{2}{l}{Parameters of the standard $\Lambda$CDM model}   \\[0.4mm]
 \multirow{2}{*}{$\theta_{\rm MC}$} & Approximate angular size of the sound \\
                        &   horizon at recombination$^{a}$\\[0.4mm] 
$\omega_{\rm b}$        &  Physical baryon density \\[0.4mm] 
$\omega_{\rm c}$        &  Physical cold dark matter density \\[0.4mm] 
$\tau$                  &  Optical depth to reionization \\[0.4mm] 
$n_{\rm s}$             &  Scalar spectral index$^{b}$ \\[0.4mm]
$A_{\rm s}$             &  Amplitude of the scalar perturbations$^{b}$\\[1.6mm]
\multicolumn{2}{l}{ Extensions to the standard model}    \\[0.4mm]
$w_0$                   &  Present-day dark energy equation of state, $w_{\rm DE}$ \\[0.4mm]
 $w_a$                  &  Time-dependence of $w_{\rm DE}$ (assuming \\
                        &  $w_{\rm DE}(a)=w_0+w_a(1-a)$)\\[0.4mm]
$\Omega_k$              &  Curvature contribution to energy density \\[0.4mm]
$\sum m_\nu$            &  Total sum of the neutrino masses \\[0.4mm]
$\gamma$                & Power-law index of the structure growth-rate\\
                        & parameter, assuming $f(z)=\Omega_{\rm m}^{\gamma}$  \\[1.6mm]
\multicolumn{2}{l}{ Derived parameters} \\[0.4mm]
$\Omega_{\rm m}$        &  Total matter density \\[0.4mm]
$\Omega_{\rm DE}$       &  Dark energy density \\[0.4mm]
$h$                     &  Dimensionless Hubble parameter \\[0.4mm]
$\sigma_{8}$            &  Linear-theory rms mass fluctuations in spheres\\
                        & of radius $8\,h^{-1}{\rm Mpc}$ \\ [0.4mm]
$S_{8}$                 &  $\sigma_8 \sqrt{\Omega_{\rm m}/0.3}$ \\ 
\hline
\end{tabular}
$^{a}$Defined as in the July 2015 version of {\sc CosmoMC}.\\ 
$^{b}$Quoted at the pivot wavenumber of $k_0= 0.05\,h{\rm Mpc}^{-1}$.
\label{tab:parameters}
\end{table}

We derive cosmological constraints from our BOSS clustering measurements following the same methodology 
as in \citet{Sanchez2014}, with small modifications.
To avoid the complication of including the covariance between our clustering measurements 
in this section we use only the information from the wedges measured in our low and high redshift bins,
and refer to these data sets as `BOSS $\xi_{3{\rm w}}$'. We use our BOSS $\xi_{3{\rm w}}$ data set in
combination with the
latest CMB temperature and polarization power spectra from the {\it Planck} satellite \citep{Planck2015},
to which we refer simply as `Planck'. We do not include CMB lensing information.
We also use the information from the joint SDSS-II and Supernova Legacy Survey (SNLS3) Light-Curve Analysis
type Ia supernovae (SN) sample \citep[JLA][]{Betoule2014}.

\begin{table} 
\centering
  \caption{
The marginalized 68\% constraints on the most relevant cosmological parameters of the extensions of the 
$\Lambda$CDM model analysed in Sections~\ref{sec:wcdm} to \ref{sec:gamma}, obtained using 
different combinations of the data sets described in Section~\ref{sec:method}.
Appendix~\ref{sec:tables} contains a complete list of the constrains obtained in each case.
}
    \begin{tabular}{@{}lcc@{}}
    \hline
 & \multirow{2}{*}{Planck+BOSS} & Planck+BOSS  \\[0.4mm]
  &                             & +SN  \\[0.4mm]
\hline
\multicolumn{3}{l}{Constant dark energy equation of state}    \\[0.6mm]
$w_{\rm DE}$            &  $-0.991_{-0.047}^{+0.062}$  &  $-0.996\pm0.042$   \\[0.4mm]
$\Omega_{\rm m}$        &  $0.308_{-0.012}^{+0.014}$  &  $0.306\pm0.011$    \\[0.4mm]
\multicolumn{3}{l}{Time-dependent dark energy equation of state}   \\[0.6mm]
$w_0$                   & $-0.73_{-0.18}^{+0.27}$  & $-0.92 \pm  0.10$   \\[0.4mm]
$w_a$                   & $-0.83_{-0.80}^{+0.58}$ & $-0.32_{-0.36}^{+0.45}$   \\[0.4mm]
$\Omega_{\rm m}$        & $0.325\pm0.020$         & $0.308\pm0.010$       \\[0.4mm]
\multicolumn{3}{l}{Non-flat models}  \\[0.6mm]
$100\Omega_k$           &  $-0.01_{-0.31}^{+0.34}$  &  $-0.07\pm0.30$    \\[0.4mm]
$\Omega_{\rm DE}$       &  $0.715\pm0.0145$         &  $0.6941 \pm 0.0079$   \\[0.4mm]
$\Omega_{\rm m}$        &  $0.288\pm0.016$          &  $0.3052_{-0.0095}^{+0.0079}$     \\[0.4mm]
\multicolumn{3}{l}{Dark energy and curvature} \\[0.6mm]
$w_{\rm DE}$            & $-0.977_{-0.070}^{+0.076}$   &  $-0.985_{-0.049}^{+0.053}$   \\[0.4mm]
$100\Omega_k$           & $ 0.16_{-0.43}^{+0.38}$      &  $0.10_{-0.39}^{+0.36}$   \\[0.4mm]
$\Omega_{\rm m}$        & $0.308\pm 0.13$              & $0.306\pm0.010$   \\
\multicolumn{3}{l}{Massive neutrinos}   \\[0.6mm]
$\sum m_\nu$/(eV)       & $ < 0.26 $ (95\% CL)       & $ < 0.25 $ (95\% CL)  \\ 
$\Omega_{\rm m}$        & $0.310_{-0.013}^{+0.009}$  &  $0.308_{-0.011}^{+0.009}$  \\
\multicolumn{3}{l}{Deviations from GR}   \\[0.6mm]
$\gamma$                & $0.609\pm0.079$               & $ < 0.610\pm0.079 $ (95\% CL)  \\ 
$\Omega_{\rm m}$        & $0.3049_{-0.0092}^{+0.0078}$  & $0.3042_{-0.0087}^{+0.0074}$  \\
\multicolumn{3}{l}{Dark energy and modified gravity}  \\[0.6mm]
$\gamma$                & $0.65_{-0.13}^{+0.10}$    &   $0.627_{-0.099}^{+0.086}$ \\[0.4mm]
$w_{\rm DE}$            &  $-1.05_{-0.08}^{+0.10}$  &   $-1.016_{-0.046}^{+0.053}$ \\
\hline
\end{tabular}
\label{tab:extra}
\end{table}

We use the July 2015 version of {\sc CosmoMC} \citep{Lewis2002}, which in turn uses {\sc CAMB} 
to compute the linear-theory CMB and matter power spectra \citep{Lewis2000}, 
modified to compute the model of non-linearities, bias and RSD described in
Section~\ref{sec:model}.
We constrain the cosmological parameters listed in Table~\ref{tab:parameters} by
directly comparing the theoretical predictions obtained for a given model 
with the galaxy clustering measurements themselves. 
Note that this approach is different from the one followed in \citet{Acacia2016}, where 
the combined growth and geometric constraints of the various BOSS clustering analyses
(including those derived in Section \ref{sec:bao}) are used as a proxy for these measurements and 
compared with the predictions from different cosmological models.
In Section~\ref{sec:lcdm} we explore the parameter space of the standard flat $\Lambda$CDM model, where the dark energy component is characterized by an equation of state parameter $w_{\rm DE} =p_{\rm DE}/\rho_{\rm DE}= -1$,
by varying the six parameters of the upper section of Table~\ref{tab:parameters}.
In Sections~\ref{sec:wcdm} to \ref{sec:gamma} we constrain 
a number of possible extensions of the $\Lambda$CDM model by allowing for variations on the
parameters presented in the middle section of Table~\ref{tab:parameters}.
We consider more general dark energy models, 
non-zero curvature, varying contributions from massive neutrinos, 
and possible deviations from general relativity. 
Table~\ref{tab:extra} summarizes the constraints on these cosmological parameters 
obtained from the combination of the Planck CMB measurements with the full shape 
of the clustering wedges from BOSS, and when this information is combined with the 
JLA SN data.
When it is not treated as a free parameter, we assume a non-zero massive neutrino component
with a total mass $\sum m_\nu=0.06\,$eV. 
For all parameter spaces we also follow the constraints on the derived
quantities listed on the final part of Table~\ref{tab:parameters}

\begin{figure}
\includegraphics[width=0.47\textwidth]{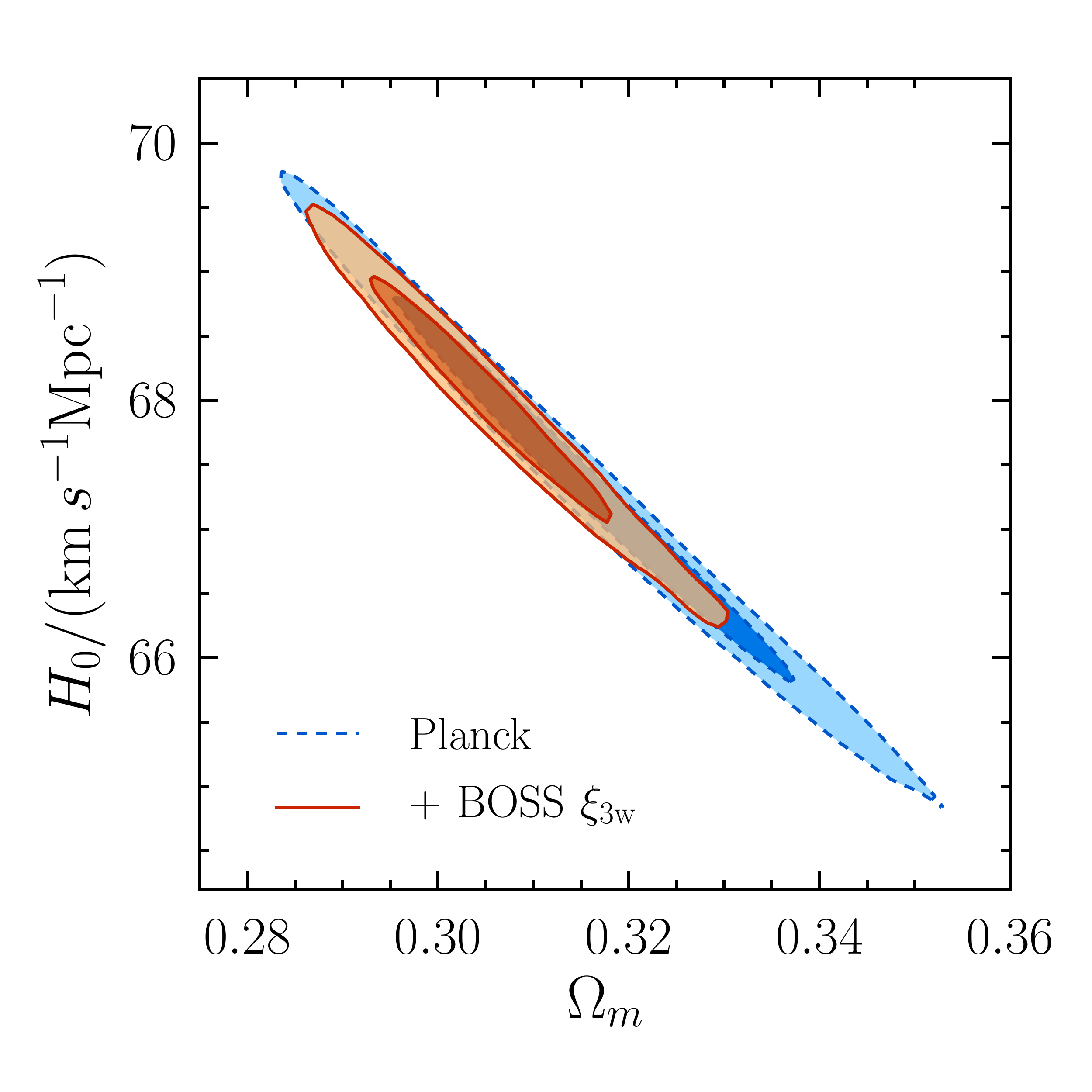}
\caption{
The marginalized posterior distribution in the $\Omega_{\rm m}$--$h$ plane for the $\Lambda$CDM 
parameter set. The dashed lines show the 68 and 95 per cent contours obtained using the CMB measurements 
from Planck alone. 
The solid contours correspond to the results obtained from the combination of the Planck data plus the 
full shape of the BOSS DR12 combined sample clustering wedges $\xi_{3{\rm w}}(s)$. 
}
\label{fig:lcdm}
\end{figure}

\subsection{The $\Lambda$CDM parameter space}
\label{sec:lcdm}

In this section we focus on the constraints on the parameters of the standard $\Lambda$CDM model. 
The dashed lines in Fig.~\ref{fig:lcdm} show the two-dimensional marginalized constraints in the 
$\Omega_{\rm m}$--$h$ plane obtained using Planck data alone. As described in \citep{Percival2002}, 
CMB-only results follow a narrow degeneracy that can be well described by a constant value of $\Omega_{\rm m}h^3$.
The solid lines in Fig.~\ref{fig:lcdm} show the result of combining the Planck data set
with the configuration space clustering wedges of BOSS. The low-redshift information provided 
by our measurements of $\xi_{3{\rm w}}(s)$ leads to a significant improvement of the obtained constraints,
with $\Omega_{\rm m}=0.3054\pm0.0087$ and $h=0.6798\pm0.0065$. These results represent constraints 
at the 2.8 and 1 per cent level and are essentially unchanged by the inclusion of the information from SN
measurements. The fact that these data sets can constrain the basic parameters of the
$\Lambda$CDM model to this precision is a clear illustration of the constraining power achieved
by current CMB and LSS measurements.
Appendix~\ref{sec:tables} gives a summary of the constraints on the full set of cosmological parameters of  
the $\Lambda$CDM model. 

The best fitting $\Lambda$CDM model gives a good description of our measurements of the clustering 
wedges, with $\chi^2$ values of 90 and 82 for the low and high-redshift bins, respectively, for 84
bins. This model is also very close to the parameters values that best describe the Planck CMB data alone, 
showing the consistency between these data sets.

\subsection{The dark energy equation of state}
\label{sec:wcdm}

\begin{figure}
\includegraphics[width=0.47\textwidth]{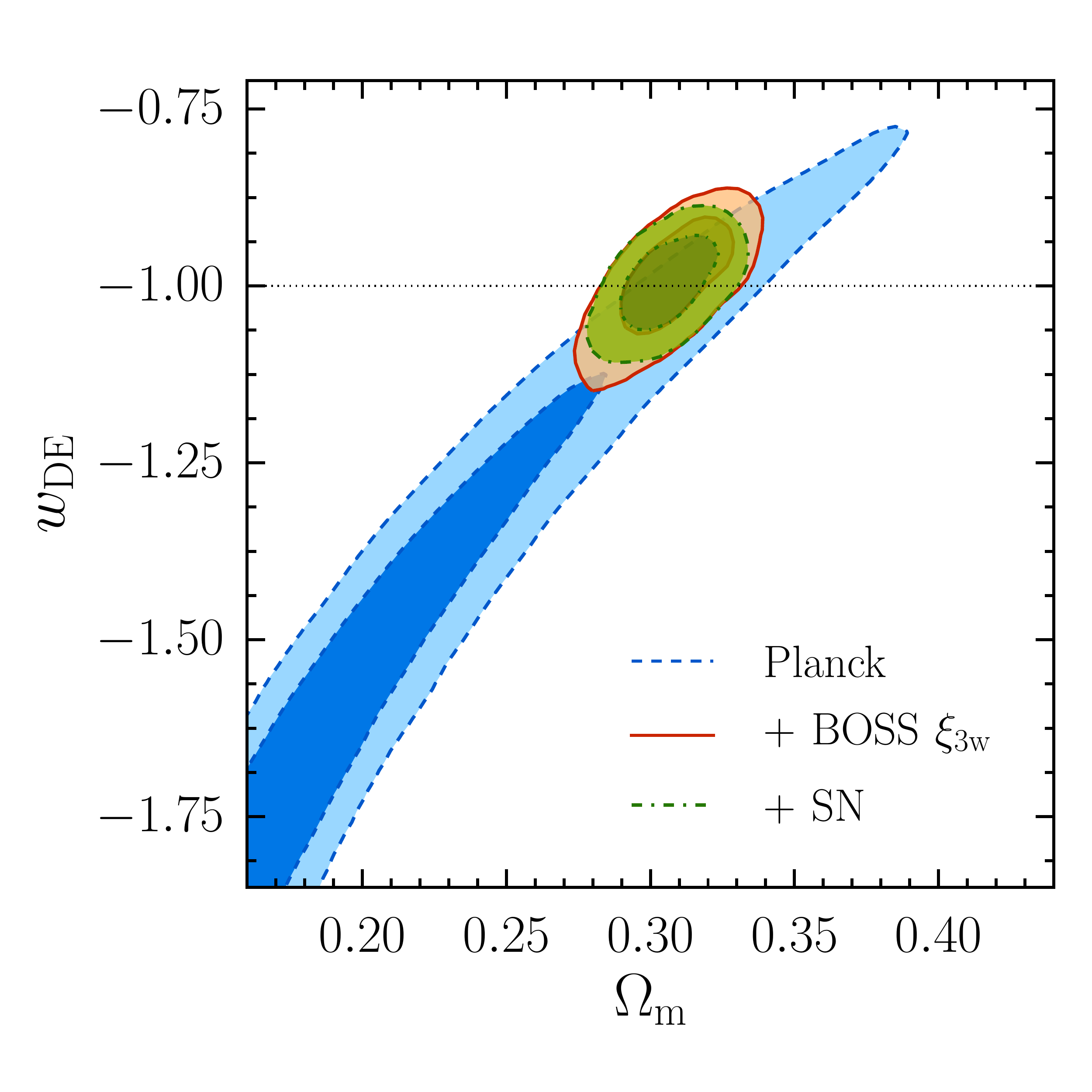}
\caption{
The marginalized posterior distribution in the $\Omega_{\rm m}$--$w_{\rm DE}$ plane for the $\Lambda$CDM 
parameter set extended by treating the redshift-independent value of $w_{\rm DE}$ as a free parameter.
The dashed lines show the 68 and 95 per cent contours obtained using Planck CMB data alone. The solid 
contours correspond to the results inferred from the combination of Planck and the BOSS
combined sample clustering wedges $\xi_{3{\rm w}}(s)$.
The dot-dashed lines indicate the results obtained when the JLA SN sample is also included in the analysis.
The dotted line indicates the standard $\Lambda$CDM value of $w_{\rm DE}=-1$.
}
\label{fig:wcdm}
\end{figure}

In the $\Lambda$CDM model the dark energy component can be described as vacuum energy, 
which behaves analogously to a cosmological constant.
In this section we explore the constraints on more general dark energy models. We start by 
treating the redshift-independent value of $w_{\rm DE}$ as an additional parameter.
The dashed lines in Fig.~\ref{fig:wcdm} correspond to the two-dimensional marginalized 
constraints in the $\Omega_{\rm m}$--$w_{\rm DE}$ plane obtained from the Planck CMB
measurements, which follow a degeneracy that spans a wide range of values of these
parameters. The solid lines in the same figure correspond to the constraints obtained when the 
Planck data is combined with the clustering wedges $\xi_{3{\rm w}}(s)$ of
the BOSS combined sample. The information encoded in these measurements provides much tighter
constraints than in the previous case, leading to $\Omega_{\rm m}=0.308_{-0.012}^{+0.014}$ and 
$w_{\rm DE}=-0.991_{-0.047}^{+0.062}$. This result is in excellent agreement with the standard
$\Lambda$CDM model value of $w_{\rm DE}=-1$, indicated by a dotted line in Fig.~\ref{fig:wcdm}.
The dot-dashed contours correspond to the results obtained by including also the information
from the JLA SN data, leading to our final constraints of 
$\Omega_{\rm m}=0.306\pm0.011$ and $w_{\rm DE}=-0.996\pm0.042$.

\begin{figure}
\includegraphics[width=0.45\textwidth]{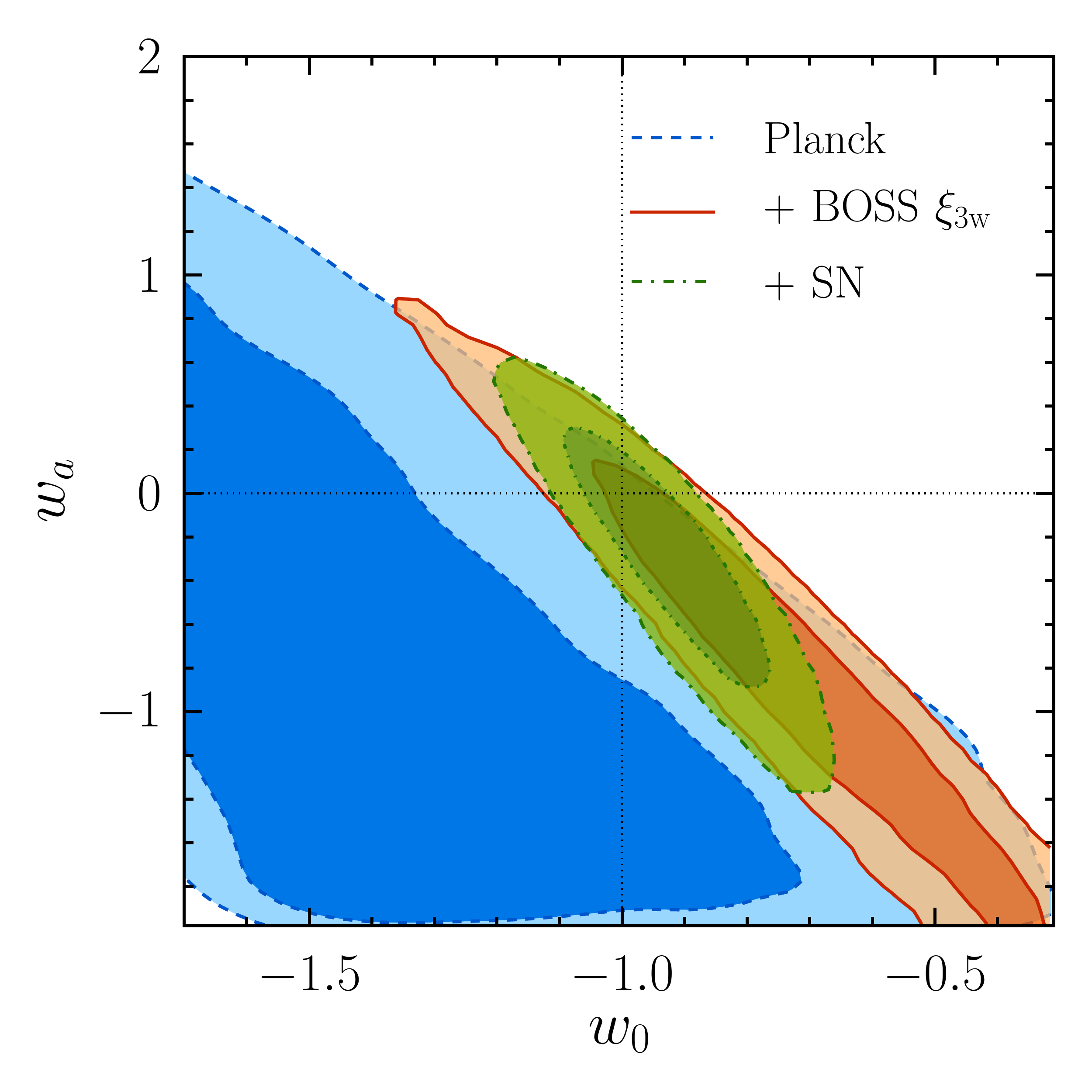}
\caption{Marginalized 68 and 95 per cent CL in the $w_0$--$w_a$ plane, the parameters controlling the 
redshift evolution of the dark energy equation of state, parametrized as in equation~(\ref{eq:wa}). 
The contours show the results obtained using the Planck CMB data alone (dashed lines), the
combination of Planck and the combined sample $\xi_{3{\rm w}}(s)$ (solid lines), and when this 
information is combined with the JLA SN data set (dot-dashed lines). The fiducial values of these 
parameters in the $\Lambda$CDM model are indicated by the dotted lines.
}
\label{fig:wacdm}
\end{figure}

In more general dark energy models the equation of state parameter might be a function of time.
To explore this possibility we use the linear parametrization of \citet{Chevallier2001} and \citet{Linder2003}
given by
\begin{equation}
w_{\rm DE}(a) = w_0 + w_a(1-a), 
\label{eq:wa}
\end{equation}
where $a$ is the scale factor and $w_0$ and $w_a$ are free parameters.
The dashed lines in Fig.~\ref{fig:wacdm} show the marginalized constraints in the $w_0$--$w_a$
plane obtained using Planck data alone, which cover a large fraction of the parameter space.
The solid lines show the effect of including the information from the BOSS $\xi_{3{\rm w}}(s)$ 
in the analysis. Although the LSS information leads to a significant reduction of the allowed 
region for these parameters, the resulting constraints on $w_0$ and $w_a$ exhibit a strong
degeneracy that allows for models whose behaviour can be significantly different to a 
cosmological constant. Additionally including information from the JLA SN sample helps to 
reduce the allowed region of the parameter space even further, leading to our final constraints 
of $w_0=-0.92 \pm 0.10$ and $w_a=-0.32_{-0.36}^{+0.45}$, in good agreement with the $\Lambda$CDM values
indicated by the dotted lines in Fig.~\ref{fig:wacdm}.

\subsection{The curvature of the Universe}
\label{sec:kcdm}

\begin{figure}
\includegraphics[width=0.47\textwidth]{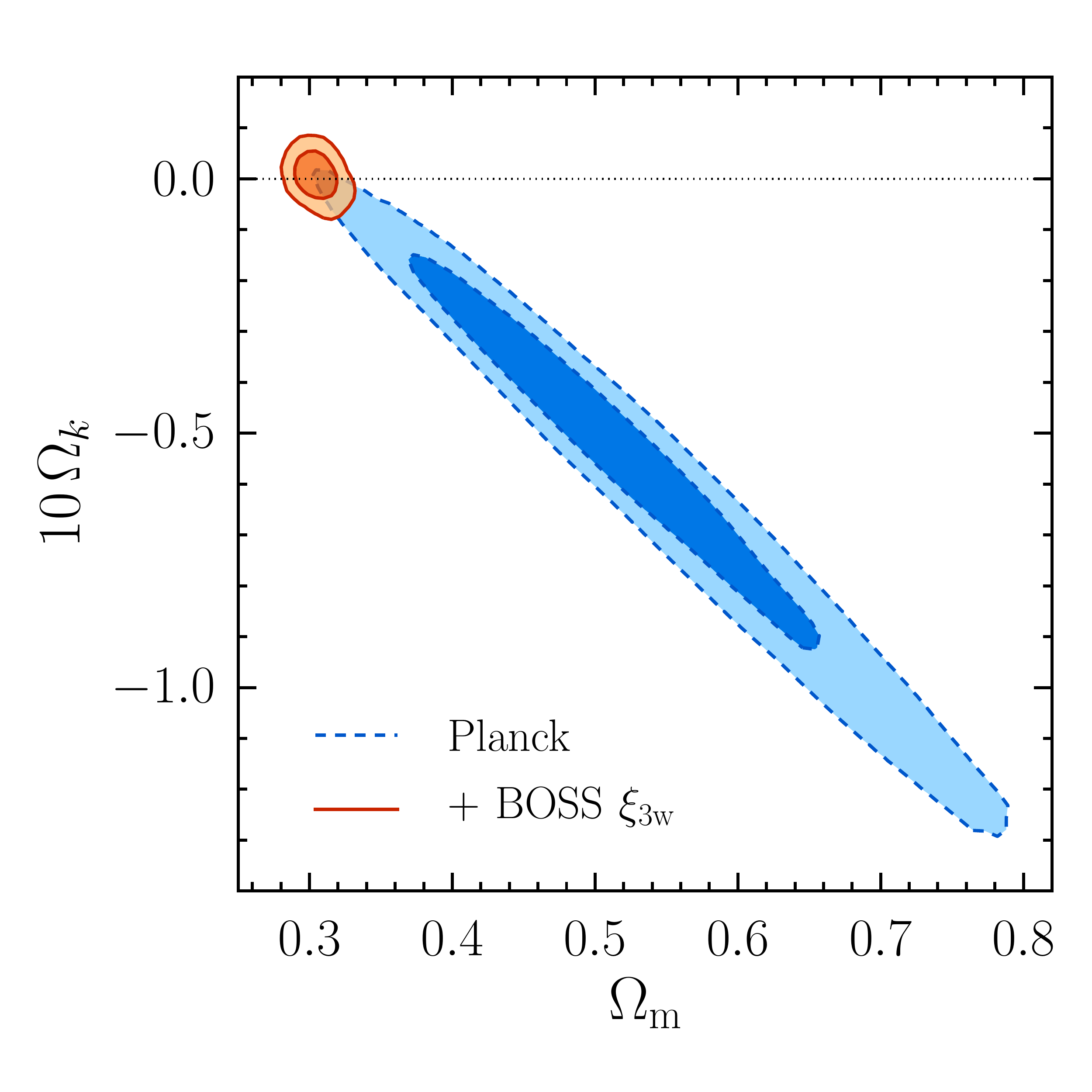}
\caption{
The marginalized posterior distribution in the $\Omega_{\rm m}$--$\Omega_{k}$ plane for the $\Lambda$CDM 
parameter set extended to allow for non-flat models.
The contours show the 68 and 95 per cent contours obtained using Planck information alone (dashed lines)
and the combination of these CMB data plus the clustering wedges of the final BOSS. 
The dotted line corresponds to the $\Lambda$CDM model, where $\Omega_{k}=0$.
}
\label{fig:klcdm}
\end{figure}

In this section we focus on non-flat models and extend the $\Lambda$CDM parameter space to models with 
$\Omega_k\neq0$. The dashed lines in Fig.~\ref{fig:klcdm} show the constraints in the $\Omega_{\rm m}$--$\Omega_{k}$
plane obtained by the Planck CMB measurements alone, which allow for significant deviations from
a flat universe due to the well-known geometric degeneracy \citep{Efstathiou1999}. 
The information from the clustering wedges from BOSS efficiently breaks this degeneracy, reducing the 
allowed region of the parameter space to a small area centred on the flat Universe value $\Omega_{k}=0$,
which is shown by the dotted line. As indicated in Table~\ref{tab:extra}, these data sets
can constrain the curvature of the Universe to 
$\Omega_{k}=-0.0001_{-0.0030}^{+0.0034}$. Additionally including the JLA SN does not significantly improve the
results over those obtained using the Planck+BOSS $\xi_{3{\rm w}}$ combination,
with a final constraint of $\Omega_{k}=-0.0007\pm 0.0030$ obtained from the combination of all datasets. 

\begin{figure}
\includegraphics[width=0.47\textwidth]{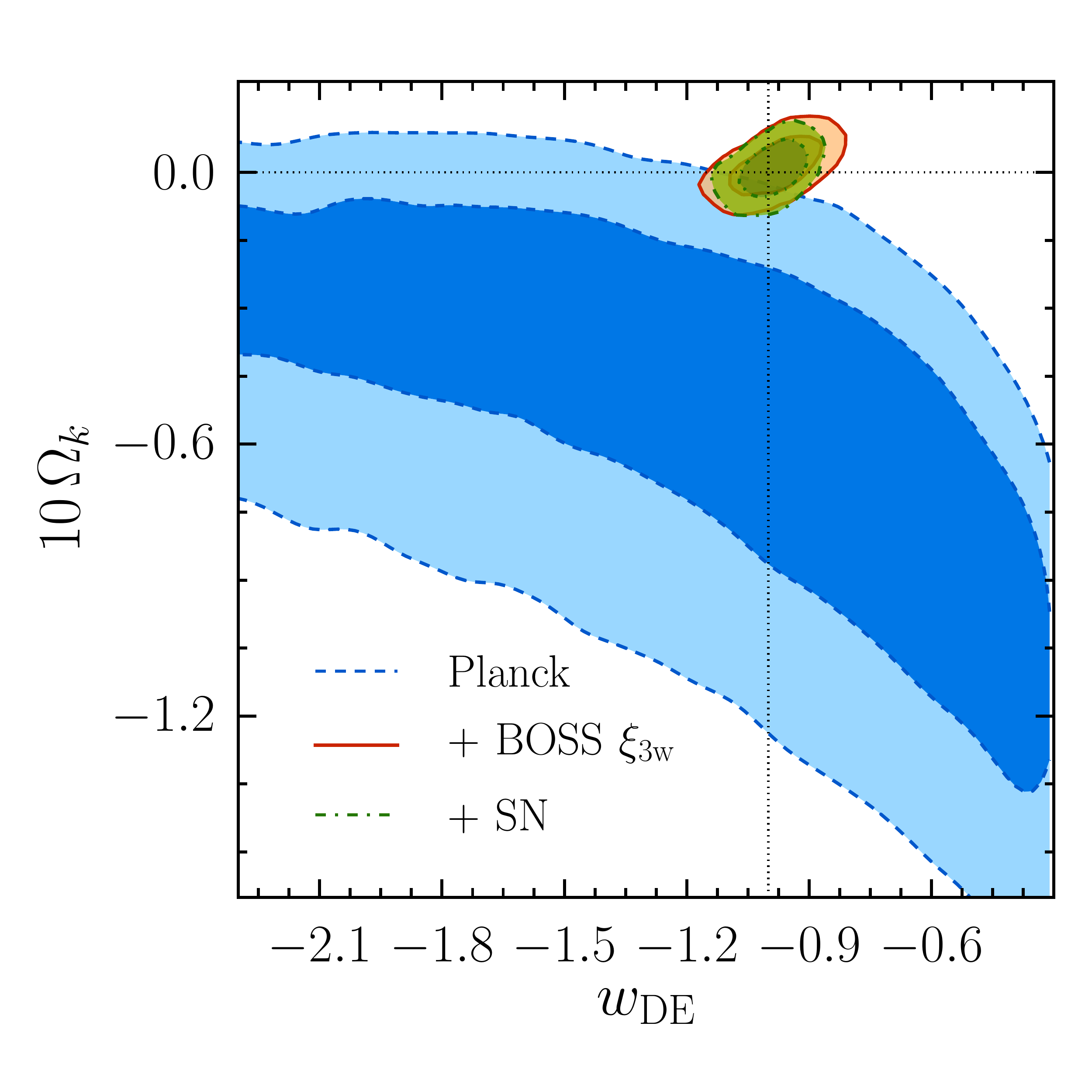}
\caption{
The marginalized constraints in the $w_{\rm DE}$--$\Omega_k$ plane for the $\Lambda$CDM 
parameter set extended by allowing for simultaneous variations on both of these parameters.
The contours correspond to the  68 and 95 per cent CL derived from the Planck CMB
data alone (dashed lines), the combination of Planck plus the clustering wedges (solid lines), and 
when the JLA SN datasets are added to the later combination (dot-dashed lines).
The dotted lines correspond to the values of these parameters in the $\Lambda$CDM model.
}
\label{fig:kwcdm}
\end{figure}

When $\Omega_k$ and $w_{\rm DE}$ are varied simultaneously, the geometric degeneracy extends to a 
two-dimensional sheet in the parameter space, degrading even more the constraints obtained from 
CMB information alone. This is shown by in the dashed contours in Fig.~\ref{fig:kwcdm}, which 
correspond to the 68 and 95 per cent CL in the $w_{\rm DE}$--$\Omega_k$ plane derived from the
Planck CMB measurements. The information in the full shape of the wedges $\xi_{3{\rm w}}(s)$
is still very effective at reducing the allowed region for these parameters, which shrinks to 
a small area around the standard $\Lambda$CDM values indicated by the dotted lines. 
In this case we find $\Omega_k=0.0016_{-0.0043}^{+0.0038}$ and $w_{\rm DE} = -0.977_{-0.070}^{+0.076}$.
As shown by the dot-dashed lines in Fig.~\ref{fig:kwcdm}, these constraints are slightly improved when 
the JLA SN information is also included in the analysis. In this case we find 
$\Omega_k=0.0010_{-0.0039}^{+0.0036}$ and $w_{\rm DE} = -0.985_{-0.048}^{+0.053}$. These constraints 
are similar to the ones we find when only one of these parameters is allowed to deviate from their 
standard values. This indicates that current constraints on the dark energy equation of state do not 
depend strongly on the assumption of a flat Universe.

\subsection{Massive neutrinos}
\label{sec:mnu}

\begin{figure}
\includegraphics[width=0.45\textwidth]{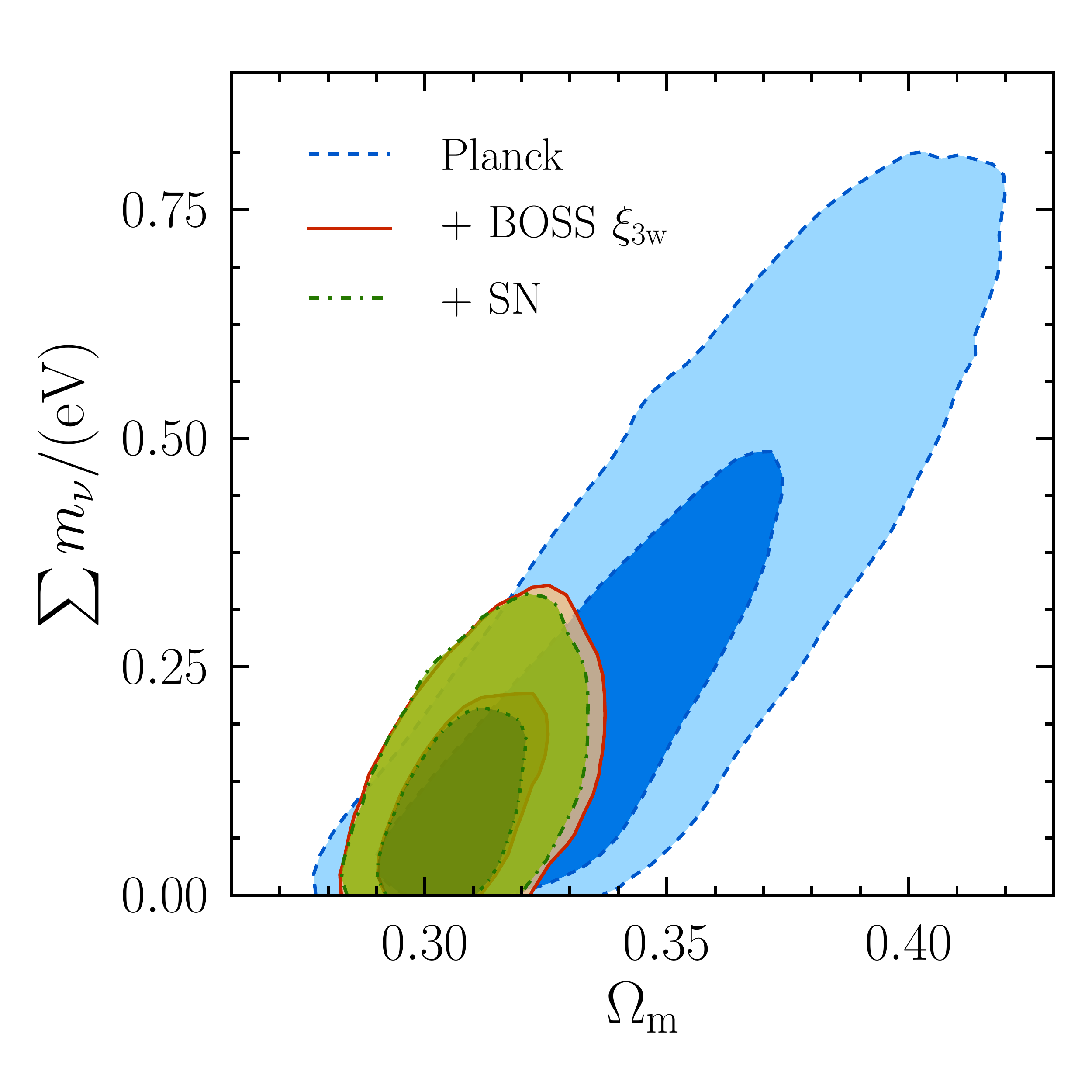}
\caption{
The marginalized posterior distribution in the $\Omega_{\rm m}$--$\Sigma m_{\nu}$ plane
for the $\Lambda$CDM parameter set extended by allowing for massive neutrinos.
The dashed and solid lines correspond to the 68 and 95 per cent CL derived
from the Planck CMB measurements alone (dashed lines) and by combining them with the 
clustering wedges $\xi_{3{\rm w}}(s)$ of the final BOSS (solid lines).
}
\label{fig:mnu}
\end{figure}

The combination of CMB and galaxy clustering measurements offers one of the best observational
windows into neutrino masses.
In the previous sections we assumed a total neutrino mass of $\sum m_{\nu} =0.06\,{\rm eV}$,
the minimum value allowed by neutrino oscillation experiments under the assumption of a normal
hierarchy \citep{Otten2008}. We now explore the constraints obtained when the total neutrino mass is allowed
to vary freely.
Fig.~\ref{fig:mnu} shows the 68 and 95 per cent constraints in the $\Omega_{\rm m}$--$\sum m_{\nu}$ plane
obtained when the $\Lambda$CDM parameter space is extended by treating $\sum m_{\nu}$ as a free
parameter. The dashed lines correspond to the results obtained using the Planck CMB data alone. 
A higher total neutrino mass leads to an increase in the redshift of matter-radiation equality,
which can be compensated by an increase in $\Omega_{\rm m}$ in order to leave the CMB power
spectrum unaffected. This is the origin of the degeneracy followed by the CMB-only constraints. 
Including the low redshift information
from the BOSS clustering wedges helps to break this degeneracy, significantly improving the
constraints. In this case we find a limit of $\sum m_{\nu} < 0.25\,{\rm eV}$ at the 95 per cent CL, 
which is almost unchanged by additionally including the JLA SN data.

\subsection{Consistency with GR}
\label{sec:gamma}

In the context of GR the redshift evolution of the structure growth-rate parameter
can be accurately computed as
\begin{equation}
f(z)=\Omega_{\rm m}(z)^{\gamma},
\label{eq:growth_gamma}
\end{equation}  
with $\gamma=0.55$ with a small correction depending on the value of $w_{\rm DE}$ \citep{Linder2007}.
This means that measurements of $f(z)$ as those obtained from anisotropic clustering measurements
can be used as a test of the predictions of GR. This information is essential to distinguish between the 
dark energy and modified gravity scenarios for the origin of the current phase of accelerated expansion 
of the Universe \citep{Zhang2007,Guzzo2008}. 
The measurements of $f(z)$ obtained from anisotropic clustering measurements could be directly compared
with the predictions of specific modified gravity models \citep[e.g.][]{2013PhRvD..88h4029W,2013MNRAS.436...89R,Taruya2014,2015PhRvD..92d3522S,Barreira2016}.
Here we follow a simpler approach and treat $\gamma$ in equation~(\ref{eq:growth_gamma}) as a free parameter. In this way, the information on the growth of structure contained in our galaxy 
clustering measurements can be used as a consistency test of GR. 
Assuming $w_{\rm DE}=-1$, a detection of a deviation from $\gamma=0.55$
can be interpreted as an indication that the growth of density fluctuations is not consistent with
the predictions of GR.

\begin{figure}
\includegraphics[width=0.43\textwidth]{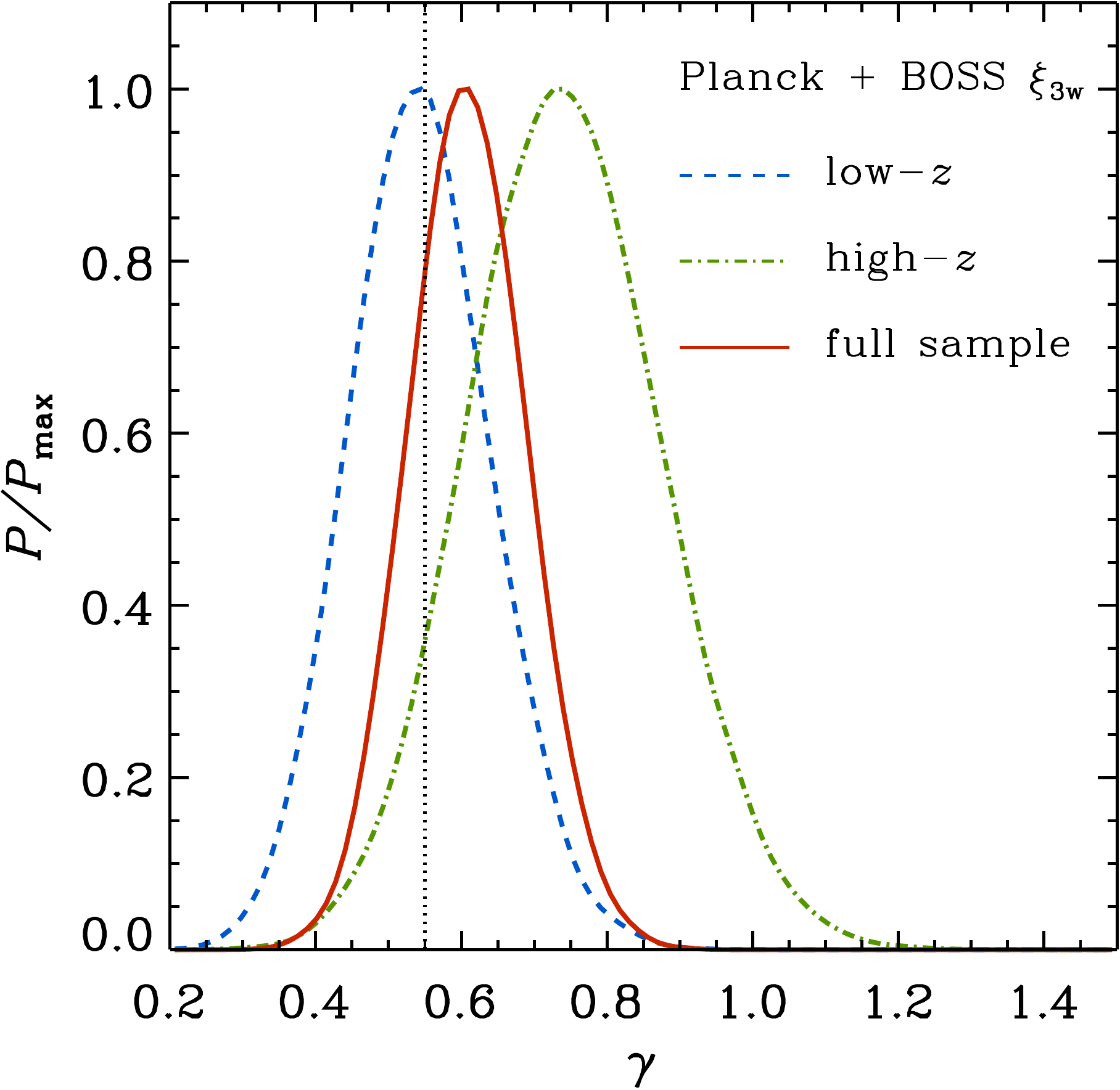}
\caption{
The one-dimensional marginalized posterior distribution of the value on the value
of the power-law index of the structure growth-rate parameter $\gamma$ derived from 
the combination of the CMB measurements from Planck and the 
clustering wedges $\xi_{3{\rm w}}(s)$ of the final BOSS sample (solid lines).
These results are consistent with the value of $\gamma=0.55$ predicted by GR, 
which is indicated by the dotted line.
}
\label{fig:gamma}
\end{figure}

We tested the consistency of our clustering measurements with GR by extending the $\Lambda$CDM parameter
space using equation~(\ref{eq:growth_gamma}) to compute $f(z)$ and treating $\gamma$ as a free parameter.
The solid line in Fig.~\ref{fig:gamma} corresponds to the one-dimensional marginalized constraints on
$\gamma$ obtained from the combination of the Planck CMB measurements with the full shape of the clustering 
wedges $\xi_{3{\rm w}}(s)$ from BOSS. In this case we find $\gamma= 0.609\pm 0.079$, in good agreement with the GR prediction of $\gamma=0.55$ indicated by the vertical dotted line. Additionally including the
JLA SN data does not improve this result.

\begin{figure}
\includegraphics[width=0.45\textwidth]{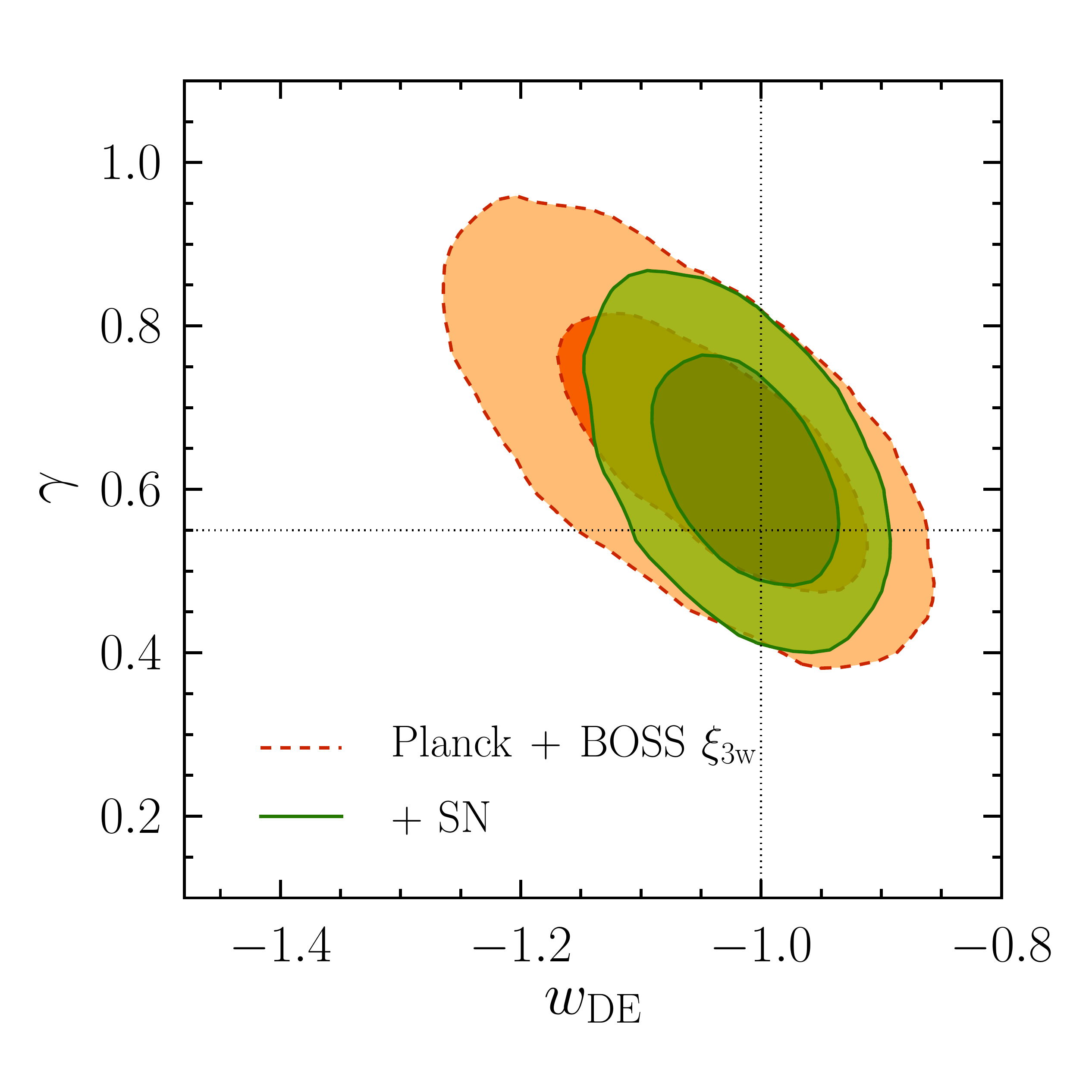}
\caption{
The marginalized posterior distribution in the $w_{\rm DE}$--$\gamma$ plane for the $\Lambda$CDM 
parameter set extended by allowing for simultaneous variations on both of these parameters.
The contours correspond to the  68 and 95 per cent CL derived from the combination of the 
Planck CMB measurements plus the clustering wedges (solid lines), and 
when the JLA SN data set is also added to the analysis (dot-dashed lines).
The dotted lines correspond to the values of these parameters in the standard $\Lambda$CDM + GR model.
}
\label{fig:gammaw}
\end{figure}

If the growth of structure is assumed to follow the predictions of GR of equation~(\ref{eq:growth_gamma})
with $\gamma=0.55$, 
the measurements of the redshift evolution of $f(z)$ obtained from RSD can be translated into constraints 
on the matter density parameter. When this assumption is relaxed by allowing $\gamma$ to vary freely this
information is lost, leading to weaker constraints on $w_{\rm DE}$ \citep{Amendola2005}.
To test this we extended the $\Lambda$CDM parameter space by allowing for simultaneous variations
of $w_{\rm DE}$ (assumed time independent) and $\gamma$.
Fig.~\ref{fig:gammaw} presents the two-dimensional
marginalized constraints in the $\gamma$--$w_{\rm DE}$ plane obtained
by means of the Planck+BOSS $\xi_{3{\rm w}}$ combination (dashed lines), and when these data are combined
with the JLA SN sample (solid lines).
Including $\gamma$ as a free parameter degrades the constraints on the dark energy equation of state
with respect to the results of Section~\ref{sec:wcdm}. 
In this case we find $w_{\rm DE}=-1.05_{-0.08}^{+0.10}$ and $\gamma=0.65_{-0.13}^{+0.10}$. 
Including the JLA SN data reduces the allowed region for these parameters, 
leading to $w_{\rm DE}= -1.016_{-0.046}^{+0.053}$  and 
$\gamma=0.627_{-0.099}^{+0.086}$, similar to the ones derived when these parameters are varied
separately and are in agreement with the standard $\Lambda$CDM+GR cosmological model.

\section{BAO and RSD constraints}
\label{sec:bao}

\begin{figure*}
\centering
\includegraphics[width=0.93\textwidth]{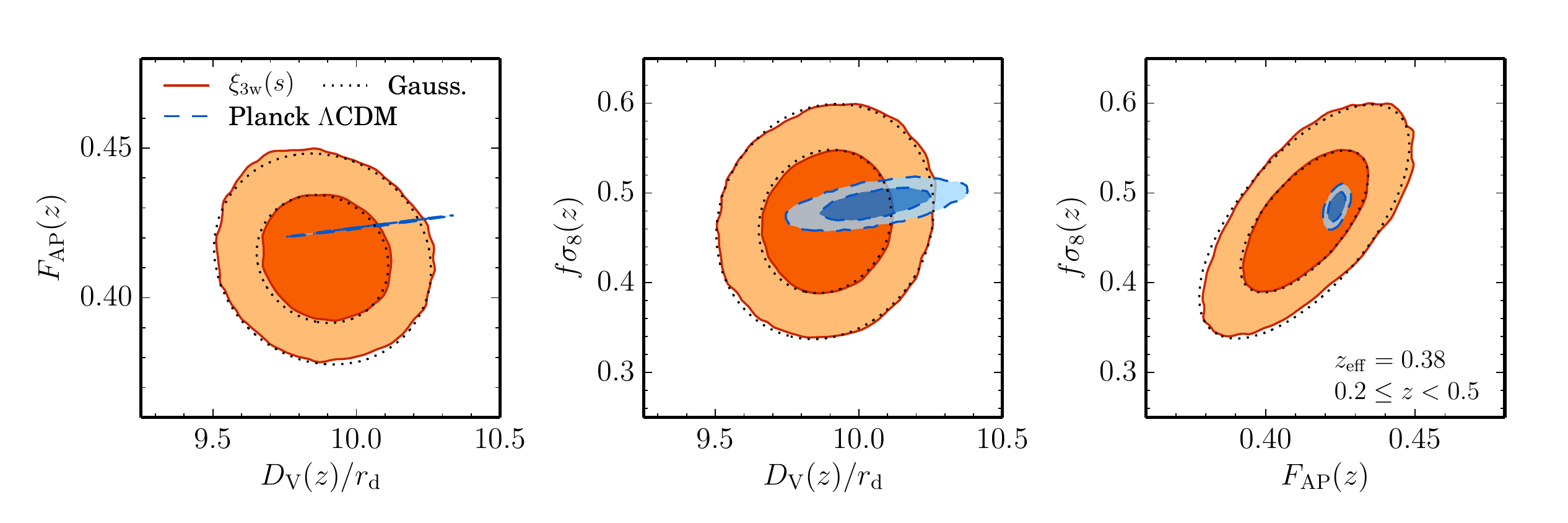}
\includegraphics[width=0.93\textwidth]{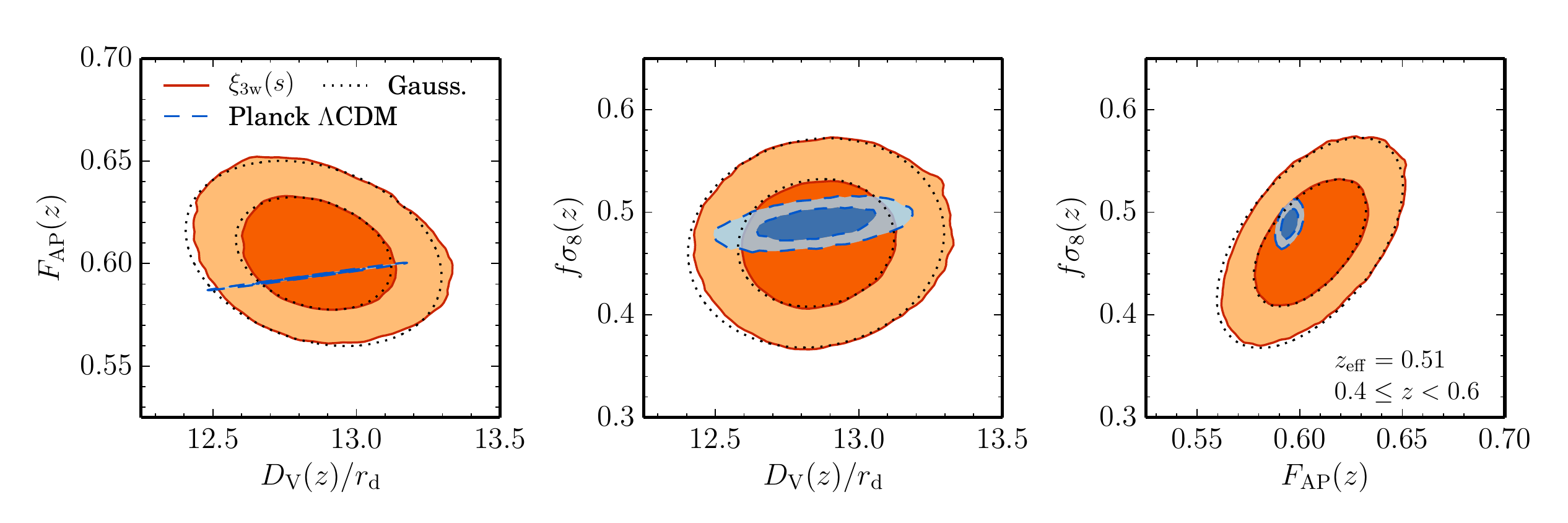}
\includegraphics[width=0.93\textwidth]{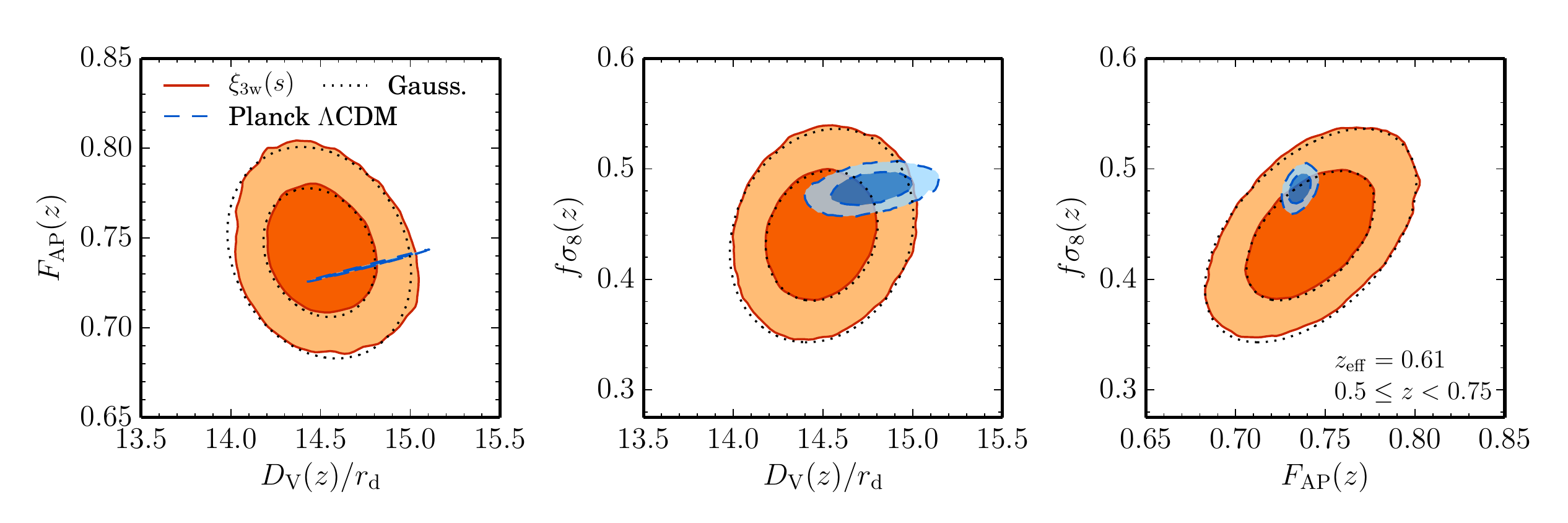}
\caption{
Two-dimensional 68 and 95 per cent marginalized constraints on $D_{\rm V}(z)/r_{\rm d}$, 
$F_{\rm AP}(z)$ and $f\sigma_8(z)$. The solid lines show the results obtained from the measurements 
of the clustering wedges $\xi_{3{\rm w}}(s)$ of the final BOSS combined sample in each of our three 
redshift bins. The dotted lines show the Gaussian approximation of these results using the mean 
values and covariance matrices of Tables~\ref{tab:results_BAO} and \ref{tab:cov_BAO}.
The dashed lines correspond to the constraints inferred from the Planck CMB measurements 
under the assumption of a $\Lambda$CDM model. 
}
\label{fig:baorsd}
\end{figure*}

In most anisotropic clustering analyses, the cosmological information contained in the full
shape of the clustering measurements is compressed into constraints on the parameter
combinations $D_{\rm M}(z)/r_{\rm d}$, 
$H(z)r_{\rm d}$ and $f\sigma_8(z)$ and their respective covariance matrix. Alternatively, these constraints
are often expressed in terms of the analogous combinations $D_{\rm V}(z)/r_{\rm d}$, where 
\begin{equation}
D_{\rm V}(z)=\left(D_{\rm M}(z)^2\frac{cz}{H(z)}\right)^{1/3},
\end{equation} 
and the Alcock-Paczynski parameter
\begin{equation}
F_{\rm AP}(z) = D_{\rm M}(z)H(z)/c.
\end{equation}
This information is then used as a proxy for the LSS measurements
when deriving constraints on cosmological parameters. 
Here we use the model described in Section~\ref{sec:model} to derive constraints on
these parameters from the clustering wedges $\xi_{3{\rm w}}$ of the final BOSS combined galaxy sample. 
To this end, we fixed the values of $\omega_{\rm b}$, $\omega_{\rm c}$ and $n_{\rm s}$ to match the 
best-fitting $\Lambda$CDM model to the CMB measurements from Planck (fixing in this way the shape of the
linear-theory power spectrum) and treated the values of $\alpha_{\perp}$, $\alpha_{\parallel}$
and $f\sigma_8$ as free parameters using separately the clustering wedges of each redshift bin. 
The nuisance parameters of the model, $b_1$, $b_2$, $\gamma_3^-$ and $a_{\rm vir}$, are also included in our
MCMC and marginalized over.
This reproduces the analysis of the Patchy mock catalogues described in Section~\ref{sec:patchy_rsd}
on the real clustering measurements from BOSS.
The lines in Fig.~\ref{fig:dr12_3w} correspond to the best-fit models obtained in this way for each of 
our redshift bins. 

The solid lines in Fig.~\ref{fig:baorsd} show the two-dimensional marginalized posterior
distributions of $D_{\rm V}(z)/r_{\rm d}$, $F_{\rm AP}(z)$ and $f\sigma_8(z)$ for each of our three redshift 
bins. The dotted lines in the same figure correspond to the Gaussian approximation of these constraints,
which give a good description of the full distributions.  
The corresponding mean values and their covariance matrices are listed in tables \ref{tab:results_BAO} and 
\ref{tab:cov_BAO}, respectively.
A comparison of our results with those of our companion papers is 
presented in \citet{Acacia2016}, where they are combined into a final set of BOSS consensus constraints
using the methodology described in \citet{Sanchez2016}.
The dashed lines in Fig.~\ref{fig:baorsd} correspond to the constraints inferred from 
the Planck CMB measurements under the assumption of a $\Lambda$CDM model. The agreement between these
results and the ones obtained from the BOSS clustering wedges indicates the consistency between these
data sets and their agreement with the $\Lambda$CDM model.

\begin{table} 
\centering
  \caption{Mean values and 68 per cent CL on $D_{\rm V}(z)/r_{\rm d}$, $F_{\rm AP}(z)$ and 
$f\sigma_8(z)$ obtained from the clustering wedges $\xi_{3{\rm w}}(s)$ of the final BOSS
combined sample in each of our three redshift bins.}
\begin{tabular}{cccc}
\hline
  Parameter                &  $z_{\rm eff}=0.38$  & $z_{\rm eff}=0.51$   & $z_{\rm eff}=0.61$    \\
\hline
$D_{\rm V}(z)/r_{\rm d}$   &  $9.89 \pm 0.15$    & $12.86 \pm 0.18$   & $14.51 \pm 0.21$       \\
$F_{\rm AP}(z)$            &  $0.413 \pm 0.014$  & $0.605 \pm 0.018$  & $0.742 \pm 0.024$       \\
$f\sigma_8(z)$             &  $0.468 \pm 0.052$  & $0.470 \pm 0.041$  & $0.439 \pm 0.039$        \\
\hline
\end{tabular}
\label{tab:results_BAO}
\end{table}

\begin{table*} 
\centering
  \caption{Covariance matrices associated with the constraints on $D_{\rm V}(z)/r_{\rm d}$, 
  $F_{\rm AP}(z)$ and $f\sigma_8(z)$ obtained from the clustering wedges of the final BOSS combined 
  sample in each of our three redshift bins.}
\begin{tabular}{cccc}
\hline
  Parameter                &  $D_{\rm V}(z)/r_{\rm d}$ & $F_{\rm AP}(z)$ & $f\sigma_8(z)$ \\
\hline
\multicolumn{3}{c}{$0.2 < z < 0.5$, $z_{\rm eff}=0.38$}   \\[0.6mm]
$D_{\rm V}(z)/r_{\rm d}$   &    $ 2.30928\times 10^{-2}$ & $-2.20148\times 10^{-4}$ &  $8.84051\times 10^{-4}$  \\
$F_{\rm AP}(z)$            &    $-2.20148\times 10^{-4}$ & $ 2.00547\times 10^{-4}$ &  $4.82676\times 10^{-4}$  \\
$f\sigma_8(z)$             &    $ 8.84051\times 10^{-4}$ & $ 4.82676\times 10^{-4}$ &  $2.76287\times 10^{-3}$  \\
\multicolumn{3}{c}{$0.4 < z < 0.6$, $z_{\rm eff}=0.51$}   \\[0.6mm]
$D_{\rm V}(z)/r_{\rm d}$   &    $3.24493\times 10^{-2}$ & $-8.31665\times 10^{-4}$ &  $7.00901\times 10^{-4}$   \\
$F_{\rm AP}(z)$            &   $-8.31665\times 10^{-4}$ &  $3.30709\times 10^{-4}$ &  $4.02845\times 10^{-4}$   \\
$f\sigma_8(z)$             &    $7.00901\times 10^{-4}$ &  $4.02845\times 10^{-4}$ &  $1.68999\times 10^{-3}$   \\
\multicolumn{3}{c}{$0.5 < z < 0.75$, $z_{\rm eff}=0.61$}   \\[0.6mm]
$D_{\rm V}(z)/r_{\rm d}$   &  $4.25331\times 10^{-2}$ &  $-9.32443\times 10^{-4}$ &  $1.31294\times 10^{-3}$    \\
$F_{\rm AP}(z)$            &  $-9.32443\times 10^{-4}$  & $5.62634\times 10^{-4}$ &  $4.60868\times 10^{-4}$    \\
$f\sigma_8(z)$             &  $1.31294\times 10^{-3}$  &  $4.60868\times 10^{-4}$ &  $1.51596\times 10^{-3}$   \\
\hline
\end{tabular}
\label{tab:cov_BAO}
\end{table*}

\section{Conclusions}
\label{sec:conclusions}

We have analysed the cosmological implications of the measurements of three clustering wedges 
$\xi_{3{\rm w}}(s)$ of the final galaxy samples from BOSS corresponding to SDSS-DR12.
We make use of the BOSS combined sample described in 
\citet{Reid2016}, containing the joint information of the LOWZ and CMASS
samples that were analysed separately in former studies, including also the 
Early regions that were previously excluded. 

We have focussed on adjusting our analysis methodology to maximize the information
extracted from the BOSS data. 
We implemented a state-of-the-art description of the effects of the non-linear
evolution of density fluctuations, bias and RSD that allowed us to extract 
information from the full shape of our clustering measurements including smaller scales than 
in previous analyses. We performed extensive tests of this model using various 
N-body simulations and BOSS mock catalogues, showing that it can be used to extract
cosmological information from our measurements of three clustering wedges for 
scales $s \gtrsim 20\,h^{-1}{\rm Mpc}$ without introducing any significant systematic errors.

We used the information from our clustering measurements in combination with the latest 
CMB measurements from Planck and the JLA SN sample to constrain the parameters of the 
$\Lambda$CDM model and a number of its potential extensions, including more general
dark energy models, non-flat universes, neutrino masses and possible deviations from 
the predictions of GR. Our results are completely consistent with the standard 
$\Lambda$CDM plus GR cosmological paradigm. When this model is extended by allowing one
additional parameter to vary freely, the combination of the CMB data from Planck
and our BOSS LSS measurements is enough to put tight constraints on the additional
variable, with the SN data leading only to marginal improvements. 
The SN information is most useful when more than one additional parameter is included
in the analysis, leading to final constraints in agreement with the canonical $\Lambda$CDM 
values. The full data set combination can constrain the dark energy equation of state 
parameter to $w_{\rm DE}=-0.996\pm0.042$ when assumed time-independent, with no indication
of a departure from this value when it is allowed to evolve according to equation~(\ref{eq:wa}).
The simultaneous variation of additional cosmological parameters does not affect
this limit significantly.
Our results are also completely consistent with the flat-Universe prediction from the most simple
inflationary models, with $\Omega_{k}=-0.0007\pm 0.0030$. 
We derive tight constraints on the total sum of neutrino masses to 
$\sum m_{\nu} < 0.25\,{\rm eV}$ at 95 per cent CL. We also test the agreement of our clustering 
measurements with the predictions of GR by assuming the parametrization of equation~(\ref{eq:growth_gamma})
for the growth-rate of cosmic structure and find $\gamma = 0.609\pm 0.079$, in 
agreement with the GR value of $\gamma=0.55$. 


The information of our clustering measurements can be compressed into constraints on
the parameter combinations $D_{\rm V}(z)/r_{\rm d}$, $F_{\rm AP}(z)$ and $f\sigma_8(z)$ 
at the mean redshifts of each of our three redshift bins with their respective 
covariance matrices. These results are in excellent agreement with the predictions of 
the best-fitting $\Lambda$CDM model to the CMB measurements from Planck, highlighting the
consistency between these data sets.
Our results are combined with those of our companion papers into a final set of consensus
constraints in \citet{Acacia2016} using the methodology described in \citet{Sanchez2016}. 

Our results show that anisotropic clustering measurements 
have become one of the most powerful available cosmological probes.
By exploiting the BAO and RSD signals imprinted in these measurements, the BOSS galaxy samples
have significantly improved our knowledge of the basic cosmological parameters. 
The application of the methodology presented here to galaxy samples from future surveys such as 
the Dark Energy Spectroscopic Instrument \citep[DESI;][]{Levi:2013gra} and the ESA space mission 
\emph{Euclid} \citep{Laureijs:2011gra} will help to push our tests of the $\Lambda$CDM paradigm to even higher accuracies. 
A joint analysis of two-point statistics with higher-order measurements such as the three-point
correlation function or the bispectrum \citep{GilMarin2015}, a detailed study of redshift-space distortions
on small scales including the impact of effects such as velocity or assembly bias \citep{Reid2014}, 
or the advancement of methods to reconstruct the underlying density field \citep{Kitaura2016b}
are strategies that could help to further increase the information extracted from LSS data sets, which will
continue shaping our understanding of cosmic history

\section*{Acknowledgements}

AGS would like to thank Ximena Mazzalay for her invaluable help in the preparation
of this manuscript.
We would like to thank Riccardo Bolze, Daniel Farrow, Jiamin Hou and
Francesco Montesano for useful discussions. 
AGS, JNG and SSA acknowledge support from the Trans-regional Collaborative Research 
Centre TR33 `The Dark Universe' of the German Research Foundation (DFG). RS was partially supported by NSF grant AST-1109432. 
CDV acknowledges financial support from the Spanish Ministry of Economy and Competitiveness (MINECO) under the 2011 and 2015 Severo Ochoa Programs SEV-2011-0187 and SEV-2015-0548, and grants AYA2013-46886 and AYA2014-58308.

Funding for SDSS-III has been provided by the Alfred P. Sloan Foundation, the Participating
Institutions, the National Science Foundation, and the U.S. Department of Energy. 

SDSS-III is managed by the Astrophysical Research Consortium for the
Participating Institutions of the SDSS-III Collaboration including the
University of Arizona,
the Brazilian Participation Group,
Brookhaven National Laboratory,
University of Cambridge,
Carnegie Mellon University,
University of Florida,
the French Participation Group,
the German Participation Group,
Harvard University,
the Instituto de Astrofisica de Canarias,
the Michigan State/Notre Dame/JINA Participation Group,
Johns Hopkins University,
Lawrence Berkeley National Laboratory,
Max Planck Institute for Astrophysics,
Max Planck Institute for Extraterrestrial Physics,
New Mexico State University,
New York University,
Ohio State University,
Pennsylvania State University,
University of Portsmouth,
Princeton University,
the Spanish Participation Group,
University of Tokyo,
University of Utah,
Vanderbilt University,
University of Virginia,
University of Washington,
and Yale University.

Based on observations obtained with Planck (http://www.esa.int/Planck), an ESA science
mission with instruments and contributions directly funded by ESA Member States, NASA, and Canada.






\appendix

\section{Details of the modelling of the two-dimensional power spectrum}
\label{app:details}

In this appendix we present a more detailed description of our model of non-linear evolution, bias and redshift-space distortions.
The  operators defined in Section~\ref{sec:bias} can be expressed in Fourier space as
\begin{align}
{\cal G}_2(\kk) &= [\dD]_{12}^{\kk}\ [ (\widehat{k}_1\cdot \widehat{k}_2)^2-1]\, \theta(\kk_1) \theta(\kk_2)\\
 &\equiv [\dD]_{12}^{\kk}\ K(\kk_1,\kk_2)\, \theta(\kk_1) \theta(\kk_2),
\end{align}
with $[\dD]_n^{\kk}\equiv \dD(\kk-\kk_{n})$, $\kk_{1\ldots n} \equiv \kk_1 + \ldots + \kk_n$ and repeated Fourier arguments are understood to be integrated over. Using this equation, the cubic operator can
be written as
\begin{align}
\Delta_3{\cal G}(\kk) & = [\dD]_{12}^{\kk}\, [ (\widehat{k}_1\cdot \widehat{k}_2)^2-1]\, (\delta(\kk_1) \delta(\kk_2) \\
       &\quad -\theta(\kk_1) \theta(\kk_2))
\label{D3G}
\end{align}
Now, since in second-order perturbation theory
\begin{equation}
\delta^{(2)}(k)-\theta^{(2)}(k)=-{2\over 7} {\cal G}_2(\kk) 
\label{dmt2}
\end{equation}
we have to leading order (and fully symmetrizing)
\begin{equation}
\begin{split}
\Delta_3{\cal G}(\kk) &=- {4\over 21} [\dD]_{123}^{\kk}\, [K(\kk_{12},\kk_3)K(\kk_1,\kk_2) \\
&\quad + K(\kk_{23},\kk_1)K(\kk_2,\kk_3) +K(\kk_{31},\kk_2)K(\kk_3,\kk_1) ] \\  
&\quad \times \delta^{(1)}(\kk_1) \delta^{(1)}(\kk_2)\delta^{(1)}(\kk_3),
\label{D3Gb}
\end{split}
\end{equation}
in terms of the linear density fluctuations. 

The galaxy auto power spectrum can be written as usual, to one-loop
\begin{equation}
\begin{split}
P_{gg}(k) & = b_1^2 P(k) + b_1b_2 P_{b_1b_2}(k)+ b_1\gamma_2 P_{b_1\gamma_2}(k) \\
          &\quad +  b_2^2\, P_{b_2b_2}(k)+ b_2\gamma_2 P_{b_2\gamma_2}(k)+ \gamma_2^2 P_{\gamma_2\gamma_2}(k) \\ 
          &\quad + \, b_1 \gamma_3^- P_{b_1 \gamma_3^-}(k) + P_{\rm noise}(k).
\end{split}
\label{Pgg}
\end{equation}
Each of these contributions are given by (in the following all powers inside integrands are linear)
\begin{align}
P_{b_1b_2}(k) & = \int 2F_2(\kk-\q,\q) P(\kk-\q)P(q) d^3q, \label{Pb1b2}\\
\begin{split}
P_{b_1\gamma_2}(k) &= P_{b_1\gamma_2}^{\rm mc}(k)  + P_{b_1\gamma_2}^{\rm prop}(k) \\ 
                   &= \int 4F_2(\kk-\q,\q) K(\kk-\q,\q)P(\kk-\q)P(q) d^3q  \\ 
                   &\quad + 8 P(k) \int G_2(\kk,\q)K(\kk-\q,\q) P(q) d^3q, \label{Pb1gamma2}
\end{split}\\
P_{b_2b_2}(k) & = {1\over 2} \int  P(\kk-\q)P(q) d^3q, \label{Pb2b2} \\
P_{b_2\gamma_2}(k)& = \int 2K(\kk-\q,\q) P(\kk-\q)P(q) d^3q, \label{Pb2g2} \\
P_{\gamma_2\gamma_2}(k) &= \int 2K(\kk-\q,\q)^2 P(\kk-\q)P(q) d^3q, \label{Pg2g2} \\
P_{b_1\gamma_3^-}(k) &= -2\,{8\over 21}P(k) \int 6 K(\kk-\q,\q)K(\kk,\q) P(q) d^3q, \label{Pb1g3m}
\end{align}
Out of these there are two terms that can be reduced to 1D integrals, they are the propagator-type integrals,
\begin{equation}
\begin{split}
P_{b_1\gamma_2}^{\rm prop}(k) &= - P(k) \int 
\Bigg[\frac{(k^2 + q^2) (33 k^4 + 14 k^2 q^2 + 33 q^4)}{42\, k^2\, q^4}  \\ 
 & + \frac{(k^2 - q^2)^2  (11 k^4 + 34 k^2 q^2 + 11 q^4)}{56\, k^3 \, q^5}  \ln \frac{(k - q)^2}{(k + q)^2} \Bigg] P(q) d^3q,
\label{Pb1gamma2p}
\end{split}
\end{equation}
and 
\begin{equation}
\begin{split}
P_{b_1\gamma_3^-}(k) &= 2\,P(k) \int \Bigg[ \frac{(k^2 + q^2) (3 k^4 - 14 k^2 q^2 + 3 q^4)}{21\, k^2\, q^4} \\
   &\quad + \frac{(k^2 - q^2)^4}{28\, k^3 \, q^5}  \ln \frac{(k - q)^2}{(k + q)^2} \Bigg] P(q) d^3q.
\label{Pb1g3m2}
\end{split}
\end{equation}
The term $P_{b2 b2}$ does not go to zero at low-k therefore we
renormalize that limit as~\citep{McD0611}
\begin{equation}
P_{b_2b_2}(k) = {1\over 2} \int  \left(1-\frac{P(q)}{P(\kk-\q)}\right)P(\kk-\q)P(q) d^3q, 
\label{Pb2b2}
\end{equation}
which now goes to zero as $k^2$. This constant low-$k$ limit enters as
an additional shot noise  
\begin{equation}
P_{\rm noise}(k) = {b_2^2\over 2} \int  P(q)^2 d^3q 
\label{Pb2b2}
\end{equation}
in practice we marginalize over shot noise for power spectrum analysis~\citep{Grieb2016}, and we can ignore shot noise renormalization for the two-point function analysis.

Similarly, we have to one-loop for the cross spectrum between galaxy fluctuations and velocity divergence that,
\begin{equation}
\begin{split}
P_{g\theta}(k) &= b_1 P_{\delta\theta}(k) + b_2 P_{b_2}(k)+ \gamma_2 P_{\gamma_2}(k) \\
                        &\quad + \gamma_3^- P_{ \gamma_3^-}(k),
\end{split}
\label{Pgm}
\end{equation}
where
\begin{equation}
P_{b_2}(k) = \int G_2(\kk-\q,\q) P(\kk-\q)P(q) d^3q,
\label{Pb1b2}
\end{equation}
\begin{align}
P_{\gamma_2}(k) & = P_{\gamma_2}^{\rm mc}(k)  + P_{\gamma_2}^{\rm prop}(k), \nonumber \\
\begin{split}
            &= \int 2 \, G_2(\kk-\q,\q) K(\kk-\q,\q)P(\kk-\q)P(q) d^3q  \\
            &\quad + 4 \, P(k) \int G_2(\kk,\q)K(\kk-\q,\q) P(q) d^3q, 
\end{split}
\label{Pb1gamma2}
\end{align}
and note that $P_{\gamma_2}^{\rm prop} = P_{b_1\gamma_2}^{\rm prop}/2$ and 
$P_{\gamma_3^-}=P_{b_1\gamma_3^-}/2$.

\section{Constraints on the $\Lambda$CDM parameter space} 
\label{sec:tables}

In this appendix we summarize the constraints on the cosmological parameters of the $\Lambda$CDM model
analysed in Section~\ref{sec:lcdm}. 
Table~\ref{tab:lcdm} list the 68\% confidence limits obtained in this parameter space. 
The upper section of the table lists the constraints on the main
parameters included in the fits, while the lower section contains the results on the parameters derived
from the first set.

\begin{table}  
\centering
  \caption{
    Marginalized 68\% constraints on the cosmological parameters of the standard $\Lambda$CDM model,
obtained using different combinations of the data sets described in Section~\ref{sec:method}. 
}
    \begin{tabular}{@{}lcc@{}}
    \hline
 & \multirow{2}{*}{Planck + BOSS $\xi_{3{\rm w}}$} & Planck + BOSS $\xi_{3{\rm w}}$  \\[0.4mm]
  &                             & + SN  \\[0.4mm]
\hline
\multicolumn{2}{l}{Main parameters} &  \\[0.3mm]
       $100\,\omega_{\rm b}$      &  $2.228\pm0.020$           &  $2.229\pm0.020$    \\[0.3mm]
        $100\,\omega_{\rm c}$     &  $11.81_{-0.16}^{+0.13}$   &  $11.80_{-0.15}^{+0.13}$      \\[0.3mm]
$10^4\times\theta_{\rm MC}$       &  $104.104\pm0.042$         &  $104.107\pm0.042$   \\[0.3mm]
       $n_{\rm s}$                &  $0.9680\pm0.0048$         &  $0.9682\pm0.0048$   \\[0.3mm]
     $\ln(10^{10}A_{\rm s})$      &  $3.078\pm0.033$           &  $3.078\pm0.033$   \\[0.3mm]
\multicolumn{2}{l}{Derived parameters} &  \\[0.3mm]
        $100\Omega_{\rm DE}$      &  $69.46_{-0.79}^{+0.95}$   &  $69.52_{-0.76}^{+0.91}$   \\[0.3mm]
        $100\Omega_{\rm m}$       &  $30.54_{-0.95}^{+0.79}$   &  $30.48_{-0.91}^{+0.76}$  \\[0.3mm]
        $h$                       &  $0.6798_{-0.0062}^{+0.0070}$ &  $   0.6803_{-0.0059}^{+0.0067}$   \\[0.3mm]
        $\sigma_{8}$              &  $0.820\pm 0.014$             &  $0.820\pm 0.014$   \\[0.3mm]
        $S_{8}$                   &  $0.827_{-0.020}^{+0.018}$    &  $0.826\pm0.018$   \\[0.3mm]
\hline
\end{tabular}
\label{tab:lcdm}
\end{table}


\bsp	
\label{lastpage}
\end{document}